\documentclass[12pt,pre,aps,preprint]{revtex4}
\usepackage{psfig}

\begin{document}

\title{Local Density Fluctuations, Hyperuniformity, and Order Metrics}

\author{Salvatore Torquato}
\affiliation{Department of Chemistry and Princeton Materials Institute,
Princeton University, Princeton, NJ 08544}
\author{Frank H. Stillinger}
\affiliation{Department of Chemistry,
Princeton University, Princeton, NJ 08544}

\date{\today}

\begin{abstract}
Questions concerning the properties
and quantification of density fluctuations in point patterns continues to 
provide many theoretical challenges. The purpose of this paper is to 
characterize 
certain fundamental aspects  of {\it local} density fluctuations associated 
with  general point patterns in any space dimension $d$. Our specific
objectives are to study
the variance in the number of points contained within a regularly-shaped
window $\Omega$ of arbitrary size, and to further illuminate our understanding of {\it hyperuniform} systems, i.e., point patterns
that do not possess infinite-wavelength fluctuations.
For large windows, hyperuniform systems are characterized by a local variance
that grows only as the surface area (rather than the volume) of the window.
We derive two formulations for the number variance: 
(i) an ensemble-average formulation, which is valid for
statistically homogeneous systems, and (ii) a volume-average
formulation, applicable to a single realization of
a general point pattern in the large-system limit.
The ensemble-average formulation (which includes both
real-space and Fourier representations) enables us to show 
that a homogeneous point pattern in a hyperuniform state 
is at a ``critical-point'' of a type with
appropriate scaling laws and critical exponents, but
one in which the {\it direct correlation function} (rather
than the pair correlation function) is long-ranged.
We also prove that the nonnegativity of the local 
number variance does not add
a new realizability condition on the pair correlation. 
The volume-average formulation is superior for certain computational
purposes, including optimization studies in which
it is desired to find the particular point pattern
with an extremal or targeted value of the variance.
We prove that the simple periodic linear array yields the global minimum value
of the average variance among all infinite one-dimensional
hyperuniform patterns. We also evaluate the variance for common 
infinite periodic lattices as well as certain
nonperiodic point patterns in one, two, and 
three dimensions for spherical windows, enabling us to rank-order
the spatial patterns. Our results suggest
that the local variance may serve as a useful order metric for
general point patterns. Contrary to the 
conjecture that the lattices associated with the densest
packing of congruent spheres have the smallest
variance regardless of the space dimension, we show that for $d=3$,
the body-centered cubic lattice has a smaller variance
than the face-centered cubic lattice. Finally, for certain
hyperuniform disordered point patterns, we evaluate the direct correlation
function, structure factor, and associated critical exponents exactly.
\end{abstract}

\maketitle

\section{Introduction}

The characterization of density fluctuations in many-particle
systems is a problem of great fundamental interest in the physical
and biological sciences. In the context of liquids, it is well known that
long-wavelength density fluctuations contain crucial thermodynamic
and structural information about the system \cite{Ha86}.
The measurement of galaxy density fluctuations is one of the most powerful 
ways to quantify and study the large-scale structure of the universe \cite{Pe93,Pi02}.
Knowledge of density fluctuations in vibrated granular media has been used 
to probe the structure and collective motions of the grains \cite{Wa96}.
Recently, the distribution of density fluctuations has been
employed to reveal the fractal nature of structures within
living cells \cite{Wa02}.

Clearly, density fluctuations that occur on some arbitrary local length scale
\cite{Ve75,Zi77,Le83,Beh93,Wa96,Tr98} provide considerably more information about the system
than only long-wavelength fluctuations.
Our main interest in this paper is to characterize certain
fundamental aspects  of {\it local} density fluctuations associated 
with general point patterns in any space dimension $d$. 
The point patterns may be thought as arising from the coordinates of the 
particles in a many-particle system, such as the molecules of a liquid, 
glass, quasicrystal, or crystal, stars of a galaxy, grains of a 
granular packing, particles of a colloidal dispersion, or trees in a
forest.

\begin{figure}[bthp]
\centerline{\psfig{file=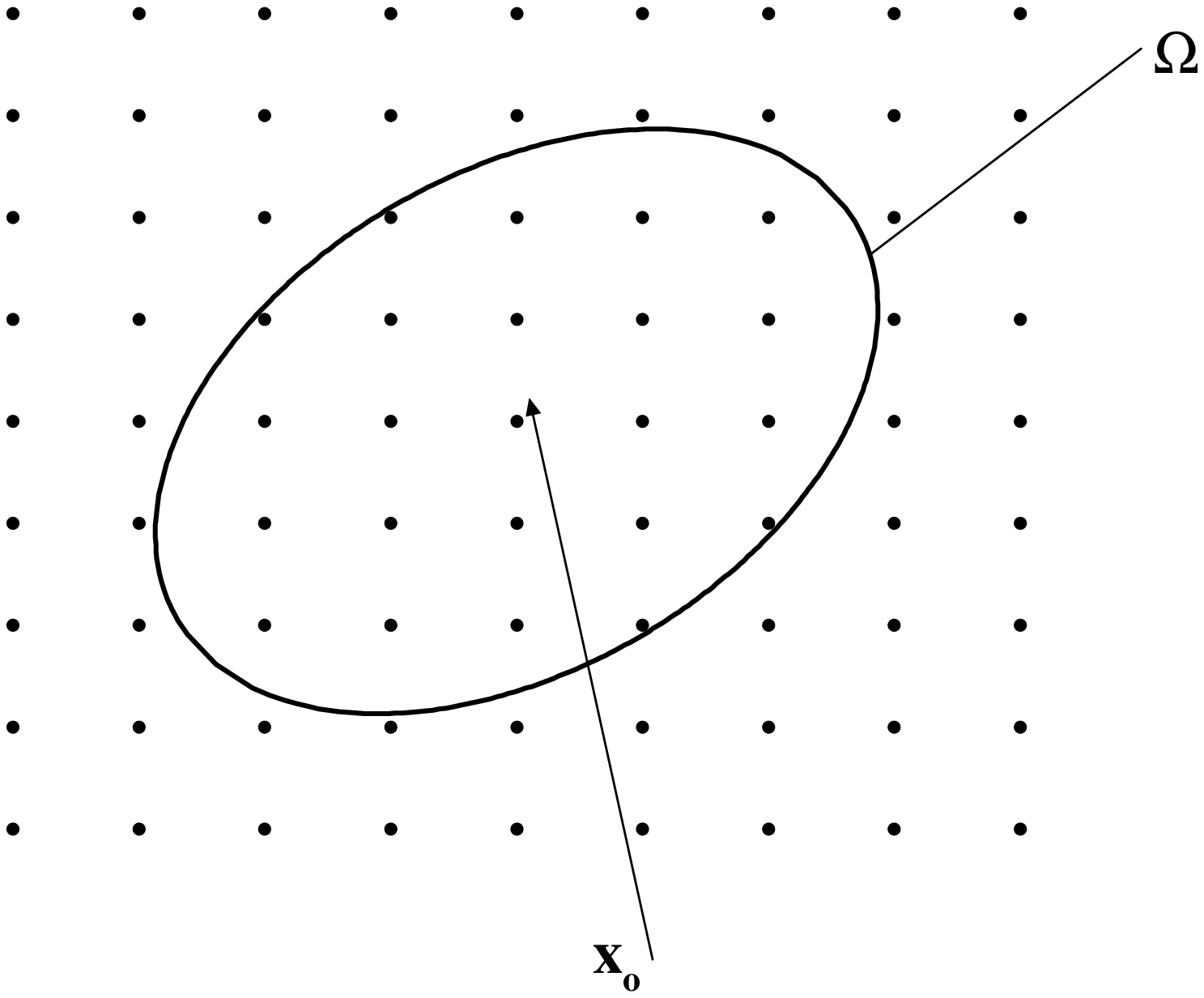,width=3.in}\psfig{file=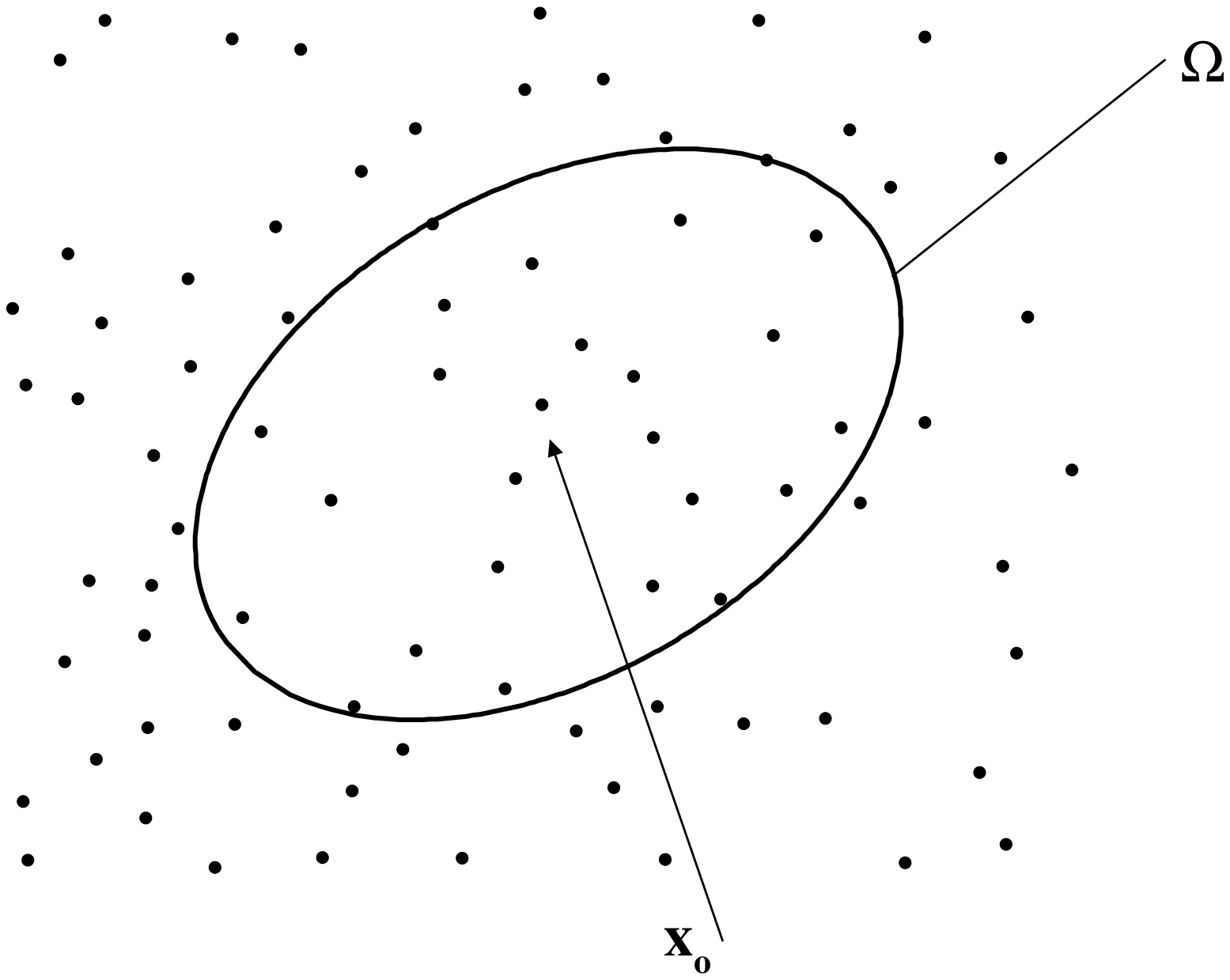,width=3.in}}
\caption{Schematics indicating a regular domain or window $\Omega$ and
its centroid $\bf x_0$ for two different point patterns. Left panel: A periodic point pattern. Right panel:
An irregular point pattern. We will show that the statistics of the
points contained within $\Omega$ for these two types of patterns
are fundamentally different from one another.}
\label{patterns}
\end{figure}

Consider an arbitrary point pattern in $d$-dimensional Euclidean space $\Re^d$.
Let $\Omega$ represent a regular domain (window) in $\Re^d$ and ${\bf x}_0$ denote
a configurational coordinate that specifies the centroid
of the window  $\Omega$. The window will  always have a fixed orientation.
There is a variety of interesting
questions that one could ask concerning the number of points contained within
$\Omega$. For example, how many points $N_\Omega$ are contained in $\Omega$
at some fixed coordinate ${\bf x}_0$? This question is a deterministic
one if the point pattern is regular and may be a statistical one
if the point pattern is irregular (see Fig. \ref{patterns}). How does the number of points
contained within some initially chosen  $\Omega$ at fixed coordinate ${\bf x}_0$
vary as the size of $\Omega$ is uniformly increased? How do the number
of points within a fixed $\Omega$ fluctuate as ${\bf x}_0$ is varied?

For a Poisson point pattern, the statistics of the number of points
contained within a regular domain are known exactly. For example,
the number variance is given by
\begin{equation}
\langle N_\Omega^2 \rangle-  \langle N_\Omega \rangle^2=  \langle N_\Omega \rangle, 
\end{equation}
where angular brackets denote an ensemble average. Letting  $\Omega$ 
be a $d$-dimensional sphere of radius $R$ and noting that $\langle N_\Omega \rangle$
is proportional to $R^d$, leads to the result that the number variance grows as the sphere
volume, i.e.,
\begin{equation}
\langle N_\Omega^2 \rangle-  \langle N_\Omega \rangle^2 \propto R^d.
\label{variance1}
\end{equation}
This result is not limited to Poisson point patterns. Indeed, a large
class of correlated irregular point patterns  obeys the variance formula
(\ref{variance1}), as we will discuss in Section II.

Can the variance grow more slowly than the volume of the domain or window?
One can show that for any statistically homogeneous
and isotropic point pattern, the variance
cannot grow more slowly than the surface area of the domain, whether it
is spherical or some other strictly convex shape \cite{Be87,Bec92}.
Thus, it is natural to ask the following question: For what class of
point pattern does the variance grow as the surface area? For a spherical
domain, we want to identify the point patterns that obey 
the variance relation for large $R$
\begin{equation}
\langle N_\Omega^2 \rangle-  \langle N_\Omega \rangle^2 \sim R^{d-1}.
\label{variance2}
\end{equation}
We will refer to such point patterns as ``hyperuniform'' systems because,
as we will see, such systems do not possess infinite-wavelength fluctuations.
(This is to be contrasted with ``hyposurficial'' systems, whose
``surface'' fluctuations vanish identically.) Additionally, it is of great interest to identify the particular point pattern
that minimizes the amplitude (coefficient) of the fluctuations
that obey (\ref{variance2}) or achieves a targeted value
of this coefficient.

\begin{figure}[bthp]
\centerline{\psfig{file=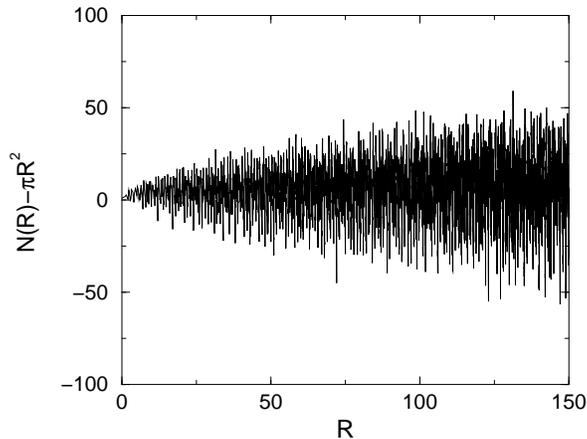,width=3.0in}}
\caption{The function $N(R)-\pi R^2$ versus $R$ for the unit-spacing
square lattice, using a circular window
of radius $R$ centered on a lattice point.}
\label{square1}
\end{figure}

Clearly, points arranged on a regular (periodic) lattice
are hyperuniform. More generally, it is
desired to know how the number of lattice points $N(R)$ contained within
a spherical window of radius $R$ varies as function of $R$ when the
sphere is centered at ${\bf x}_0$. For simplicity, let us
consider this question in two dimensions for points
arranged on the square lattice and let the center of the circular
window of radius $R$ be positioned at a point $(a_1,a_2)$ in the unit square.
The answer to this query amounts to finding all of the integer solutions of
\begin{equation}
(n_1-a_1)^2+(n_2-a_2)^2 \le R^2,
\end{equation}
a problem of interest in number theory \cite{Ken48,Ken53}.
This problem is directly related to the determination  of the
number of energy levels less than some fixed energy 
in integrable quantum systems \cite{Beh93}. It is clear that 
$N(R)$ asymptotically approaches the window area $\pi R^2$
and unit density, for large $R$.
The apparent ``random'' nature of $N(R)$ is beautifully
illustrated in Figure \ref{square1}, which shows how the
function $N(R)-\pi R^2$
grows with $R$.

It is considerably more challenging to identify non-periodic point patterns,
such as disordered and quasiperiodic ones, that are
hyperuniform. The mathematical conditions that statistically homogeneous
hyperuniform systems must obey (derived in Section II) are a necessary starting
point in identifying such hyperuniform point patterns. These conditions,
which include the counterintuitive property of a long-ranged
``direct'' correlation function, 
are determined from a general formula for the number variance
of such systems, which is obtained in Section II. 
The fact that the direct correlation
function of a hyperuniform pattern is long-ranged
is reminiscent of the behavior of the pair correlation
function of a thermal system near its critical point.
Indeed, we show that a statistically homogeneous point pattern in a hyperuniform state 
is at a ``critical-point'' of a type with
appropriate scaling laws and critical exponents. By deriving a Fourier
representation of the local variance, it is also shown
that the nonnegativity of the variance does not add
a new realizability condition on the pair correlation function beyond the known
ones. 

To date, only a few statistically homogeneous and isotropic
patterns have been rigorously shown to be 
hyperuniform. One of the aims of this paper will be to identify
other such hyperuniform examples, and to describe a procedure 
to find them systematically. This requires a formulation 
for the local variance that can be applied to a single realization of any pattern, which
is accomplished in Section III. In Section IV  we prove that the
simple periodic linear array yields the global minimum value
of the average variance  among all infinite one-dimensional
hyperuniform patterns. Interestingly, we also show that
the variance for large spherical windows enables us to rank-order common regular lattice 
and certain disordered point patterns in one,
two, and three dimensions (see Sections IV and V). 
Our results suggest that the local variance may provide 
a useful order metric for general point patterns (see Section VI). Contrary to the 
conjecture 
that the Bravais lattice associated with the densest
packing of congruent spheres has the smallest
variance regardless of the space dimension, we show that for $d=3$,
the body-centered cubic lattice has a smaller variance
than the face-centered cubic lattice. In Section V,
we evaluate the direct correlation
function, structure factor, and associated critical exponents exactly
for certain hyperuniform disordered point patterns. 
Three appendices provide analytical formulas for
key geometrical quantities required for the theory,
an evaluation of the variance  for
hard rods in equilibrium for large windows, and 
a discussion of a certain property of hyposurficial point patterns.

\section{Local Variance Formula for Realizations
of Statistically Homogeneous Systems}

A general expression for
the local number variance 
for realizations of statistically homogeneous 
point patterns in $d$ dimensions is derived. This is necessarily an
ensemble-average formulation. We obtain both a real-space
and Fourier representation of the variance.
From these results, we 
obtain formulas  for asymptotically large
windows. We show that a hyperuniform point pattern
is at a type of ``critical-point''  with
appropriate scaling laws and critical exponents, but
one in which the direct correlation function is long-ranged.

\subsection{Preliminaries}

Consider $N$ points with configuration
${\bf r}^N \equiv {\bf r}_1,{\bf r}_2,\ldots,{\bf r}_N$ in a volume $V$. 
 The local number density
 at position $\bf x$ is given by
 \begin{equation}
 n({\bf x})=\sum_{i=1}^N \delta({\bf x}-{\bf r}_i),
 \end{equation}
 where $\delta(\bf x)$ is the Dirac delta function.
The point pattern is statistically characterized by the {\it specific}
probability density function $P_N({\bf r}^N)$, where $P_N({\bf r}^N)d{\bf
r}^N$ gives the probability of finding point 1 in volume element
$d{\bf r}_1$ about ${\bf r}_1$, point 2 in volume element
$d{\bf r}_2$ about ${\bf r}_2$, $\ldots$, point $N$  in volume element
$d{\bf r}_N$ about ${\bf r}_N$. Thus, $P_N({\bf r}^N)$ normalizes
to unity and $d{\bf r}^N \equiv d{\bf r}_1,d{\bf
r}_2,\ldots,d{\bf r}_N$ represents the $Nd$-dimensional volume element. 
The ensemble average of any function $f({\bf r}^N)$ that depends
on the configuration of points is given by
\begin{equation}
\langle f({\bf r}^N) \rangle = \int_V \int_V \cdots \int_V f({\bf r}^N)
P_N({\bf r}^N)d{\bf r}^N.
\end{equation}

Because complete statistical information is usually
not available, it is convenient to introduce the reduced {\it generic}
density  function $\rho_n({\bf r}^n)$ ($n <N$), defined as
\begin{eqnarray}
\rho_n({\bf r}^n) = \frac{N!}{(N-n)!}  \int_V \cdots \int_V P_N({\bf r}^N) d{\bf r}^{N-n},
\label{rhon}
\end{eqnarray}
where $d{\bf r}^{N-n}\equiv d{\bf r}_{n+1} d{\bf r}_{n+2} \cdots  d{\bf r}_N $.
In words, $\rho_n({\bf r}^n)d{\bf r}^n$ is proportional to
the probability of finding {\it any} $n$ particles ($n \le N$)
with configuration $\bf r^n$ in volume element $d{\bf r}^n$.
In light of its probabilistic nature, it is clear that  $\rho_n({\bf r}^n)$ 
is a nonnegative quantity, i.e., $\rho_n({\bf r}^n) \ge
0,  \forall {\bf r}^n$.

For statistically homogeneous media, $\rho_{n}({\bf r}^n)$
is translationally invariant and hence depends only on the relative
displacements, say with respect to ${\bf r}_1$:
\begin{equation}
\rho_{n}({\bf r}^n)=\rho_{n}({\bf r}_{12},{\bf r}_{13},\ldots,{\bf r}_{1n}),
\end{equation}
where ${\bf r}_{ij}={\bf r}_j-{\bf r}_i$. In particular, the one-particle
function $\rho_1$ is just equal to the constant {\it number density}
\index{number density} of particles $\rho$, i.e.,
\begin{equation}
\rho_1({\bf r}_1) = \rho \equiv \lim_{ N,V\rightarrow\infty} \frac{N}{V} .
\label{thermolimit}
\end{equation}
The limit indicated in (\ref{thermolimit}) is referred to as the
{\it thermodynamic limit}. 
Since our interest
in this section is in statistically homogeneous point patterns, we now
take the thermodynamic limit. 
It is convenient to define the so-called
{\it $n$-particle correlation function}
\begin{equation}
g_n({\bf r}^n) = \frac{\rho_n({\bf r}^n)}{\rho^n}.
\label{nbody}
\end{equation}
In systems without long-range order and in which the
particles are mutually far from one another (i.e.,  ${r}_{ij}=|{\bf r}_{ij}|
\rightarrow\infty$,
$1\leq i < j \leq N$), $\rho_n({\bf r}^n) \rightarrow \rho^n$ and we have
from (\ref{nbody}) that
$g_n({\bf r}^n) \rightarrow 1$.
Thus, the deviation of $g_n$ from unity  provides a
measure of the degree of spatial correlation
between the particles, with unity corresponding to no spatial correlation.

The important two-particle quantity
\begin{equation}
g_2({\bf r}_{12}) = \frac{\rho_2({\bf r}_{12})}{\rho^2}
\label{g2-rho2}
\end{equation}
is usually referred to as the {\it pair correlation function}.
The {\it total correlation function} $h({\bf r}_{12})$ is defined
as
\begin{equation}
h({\bf r}_{12})=g_2({\bf r}_{12})-1,
\label{total}
\end{equation}
and thus is a function that is zero when there are no spatial
correlations in the system.
When the system is both statistically homogeneous 
and isotropic, the pair correlation function depends on  the radial distance
$r_{12}$ only, i.e.,
\begin{equation}
g_2({\bf r}_{12}) = g_2(r_{12}),
\end{equation}
and is referred to as the radial distribution function.
From (\ref{g2-rho2}), we see that
$\rho s_1(r) g_2(r) dr$ is proportional to the {\it
conditional
probability} of finding a particle center in a spherical shell of volume
$s_{1}(r)dr$, given that there is another at the
origin. Here $s_{1}(r)$ is the surface area of a $d$-dimensional sphere of
radius $r$, which is 
given by 
\begin{equation}
s_1(r)  =  \frac{2\pi^{d/2}r^{d-1}}{\Gamma(d/2)},
\label{area-sph}
\end{equation}
where $\Gamma(x)$ is the gamma function. Hence, for a finite system, integrating
$(N-1) g_2(r)/V$ over the volume yields $N-1$, i.e., all the particles
except the one at the origin.

Observe that the structure factor $S(\bf k)$ is related to the Fourier
transform of $h(\bf r)$, denoted by ${\tilde h}({\bf k})$, via the expression
\begin{equation}
S({\bf k})=1+\rho {\tilde h}({\bf k}).
\label{factor}
\end{equation}
The Fourier transform of some absolutely
integrable function
$f({\bf r})$ in $d$ dimensions is given by
\begin{equation}
{\tilde f}({\bf k}) = \int  f({\bf r}) e^{\displaystyle -i{\bf k}\cdot{\bf r}}\, d{\bf r},
\label{fourier}
\end{equation}
and the associated inverse operation is defined by
\begin{equation}
f({\bf r}) = \frac{1}{(2\pi)^d} \int{\tilde f}({\bf k})
e^{\displaystyle i{\bf k}\cdot{\bf r}} \, d{\bf k},
\label{fourier-inverse}
\end{equation}
where $\bf k$ is the wave vector. It is well known that
the structure factor is proportional to the scattered intensity
of radiation from a system of points and thus is obtainable
from a scattering experiment.
An important property of the structure factor is that it must be nonnegative
for all $\bf k$, i.e.,
\begin{equation}
S({\bf k}) \ge 0 \qquad \forall {\bf k}.
\label{S-k}
\end{equation}

\subsection{General Variance Formulas}

Let $\bf R$ symbolize the parameters that 
characterize the geometry of the window $\Omega$. For example, in
the case of an ellipsoidal window, $\bf R$ would represent the semi-axes of the ellipsoid. 
Let us introduce
the window indicator function 
\begin{equation}
w({\bf x}-{\bf x}_0;{\bf R})=\Bigg\{{1, \quad {\bf x} \in \Omega,
\atop{0, \quad {\bf x} \notin \Omega,}}
\label{window}
\end{equation}
for a window with a configurational coordinate ${\bf x}_0$.
The number of points $N_\Omega$ within the window at ${\bf x}_0$,
which we henceforth denote by $N({\bf x}_0;{\bf R})$,
is given by
\begin{eqnarray}
N({\bf x}_0;{\bf R}) &=& \int_V n({\bf x})
w({\bf x}-{\bf x}_0;{\bf R}) d{\bf x} \nonumber \\
&=& \sum_{i=1}^N \int_V \delta({\bf x}-{\bf r}_i) w({\bf x}-{\bf x}_0;{\bf R}) d{\bf x} \nonumber \\
&=& \sum_{i=1}^N w({\bf r}_i-{\bf x}_0;{\bf R}).
\label{N}
\end{eqnarray}
Therefore, the average number of points contained within the window
in a realization of the ensemble is
\begin{eqnarray}
\langle N({\bf R}) \rangle &=& \int_V \sum_{i=1}^N w({\bf r}_i-{\bf
x}_0;{\bf R}) P_N({\bf r}^N) d{\bf r}^N \nonumber \\
&=& \int_V \rho_1({\bf r}_1) w({\bf r}_1-{\bf x}_0;{\bf R}) d{\bf r}_1 \nonumber
\\
&=&  \rho\int_{\Re^d}  w({\bf r};{\bf R}) d{\bf r} \nonumber \\
&=& \rho v_1({\bf R}),
\label{N(R)}
\end{eqnarray}
where $v_1({\bf R})$ is the volume of
a window with geometric parameters  $\bf R$.
Note that translational invariance of the point pattern, invoked
in the third line of relation (\ref{N(R)}), renders the average
$\langle N({\bf R}) \rangle$ independent of the window coordinate 
${\bf x}_0$.

Similarly, ensemble averaging the square of (\ref{N}) and using
relation (\ref{N(R)}) gives the local number variance as
\begin{eqnarray}
\langle N^2({\bf R}) \rangle-  \langle N({\bf R}) \rangle^2 &=&
\int_V \rho_1({\bf r}_1) w({\bf r}_1-{\bf x}_0;{\bf R}) d{\bf r}_1 \nonumber \\
&+& \int_V \int_V[\rho_2({\bf r}_1,{\bf r}_2) - \rho_1({\bf r}_1)\rho_1({\bf r}_2)]
w({\bf r}_1-{\bf x}_0;{\bf R}) 
w({\bf r}_2-{\bf x}_0;{\bf R}) d{\bf r}_1 d{\bf r}_2 \nonumber \\
&=& \langle N({\bf R}) \rangle \Bigg[ 1+\rho\int_{\Re^d}  h({\bf r})
\alpha({\bf r};{\bf R}) d{\bf r}\Bigg],
\label{N2}
\end{eqnarray}
where $h({\bf r})$ is the total correlation function defined by (\ref{total}),
\begin{equation}
\alpha({\bf r};{\bf R})= \frac{v_2^{\mbox{\scriptsize int}}({\bf r};
{\bf R})}{v_1({\bf R})},
\label{I}
\end{equation}
and
\begin{equation}
v_2^{\mbox{\scriptsize int}}({\bf r};{\bf R})= \int_{\Re^d} w({\bf r}_1-{\bf x}_0;{\bf R})
w({\bf r}_2-{\bf x}_0;{\bf R}) d{\bf x}_0
\end{equation}
is the intersection volume of two windows (with the same orientations)
whose centroids  are separated by the displacement
vector  ${\bf r} ={\bf r}_1-{\bf r}_2$ \cite{To02a}. 
Appendix A provides explicit analytical formulas for
the intersection volume for spherical windows in arbitrary
dimension $d$. As before,
statistical homogeneity, invoked in the second line of (\ref{N2}),
renders the variance independent of ${\bf x}_0$.

\bigskip

\noindent{\it Remarks:}
\smallskip

\noindent{1. Formula (\ref{N2}) was previously derived by Landau and Lifschitz \cite{La},
although they did not explicitly indicate the scaled intersection
volume function $\alpha({\bf r};{\bf R})$. Martin and Yalcin \cite{Ma80}
derived the analogous formula for charge fluctuations in classical
Coulombic systems.}
\smallskip

\noindent{2. The local variance formula (\ref{N2}) is closely related
to one associated with the local volume fraction fluctuations
in two-phase random heterogeneous materials \cite{To02a,Lu90}. Both formulas
involve the scaled intersection volume function $\alpha({\bf r};{\bf R})$.
The essential difference
is that the variance for local volume fraction fluctuations
involves a different correlation function from $h(\bf r)$, namely, 
the probability of finding two points, separated
by a displacement $\bf r$, both in the same phase.}
\smallskip

\noindent{3. The existence of the integral in (\ref{N2}) requires
that the product $h({\bf r})\alpha({\bf r};{\bf R})$ be integrable.
For finite size windows, this will be the case
for bounded $h(\bf r)$ because $\alpha({\bf r};{\bf R})$
is zero beyond a finite distance. For infinitely
large windows, $\alpha({\bf r};{\bf R})=1$, and integrability
requires that $h(\bf r)$ decays faster than $|{\bf r}|^{-d+\epsilon}$
for some $\epsilon >0$. For systems in thermal equilibrium, this
will be the case for pure phases away from critical points.
The structure factor S({\bf k})  [defined by (\ref{factor})] at ${\bf k}=\bf 0$ diverges as a  
thermal critical point
is approached, implying that $h({\bf r})$ becomes long-ranged, i.e., 
decays slower than $|{\bf r}|^{-d}$ \cite{Hu87}.}
\bigskip

An outstanding question in statistical physics is: What are the existence
conditions for a valid (i.e., physically
realizable) total correlation function $h({\bf r})$ \cite{To02b}
of a point process at fixed density $\rho$?
The generalization of
the Wiener--Khinchtine theorem for multidimensional spatial stochastic
processes \cite{Cr93} states
a necessary and sufficient condition for
the existence of  an autocovariance function 
of a general stochastically continuous homogeneous process  
is that it has a spectral (Fourier--Stieltjes) representation
with a nonnegative bounded measure. If
the autocovariance is absolutely integrable, this implies
that its Fourier transform must be nonnegative.
The total correlation function $h(\bf r)$ 
is the nontrivial part of the autocovariance function for a point
process, i.e., it excludes the delta function at the origin.
The fact that $h(\bf r)$ comes from a statistically homogeneous
point process, however, would further restrict the existence conditions
on $h(\bf r)$ beyond the Wiener--Khinchtine condition, which
amounts to the nonnegativity of the structure factor.
Obviously, besides the condition that $S(\bf k) \ge 0$, we have
the pointwise condition $h({\bf r}) \ge -1$ for all $\bf r$.
The determination of other realizability conditions on $h(\bf r)$
is a open question \cite{To02b}.

Thus, it is interesting to inquire whether the nonnegativity of the 
local number variance, given
by formula (\ref{N2}), is a new condition on $h(\bf r)$ beyond the
nonnegativity of the structure factor $S({\bf k})$.
As we now prove, the answer is no. By Parseval's
theorem for Fourier transforms \cite{Sn95}, we can rewrite the general variance formula
(\ref{N2}) for an arbitrarily shaped (regular) window as
\begin{equation}
 \langle N^2({\bf R}) \rangle-  \langle N({\bf R}) \rangle^2=
\langle N({\bf R}) \rangle \Bigg[1+\frac{\rho}{(2\pi)^d} \int
{\tilde h}({\bf k}){\tilde \alpha}({\bf k};{\bf R}) d{\bf k}\Bigg],
\end{equation}
where
\begin{equation}
{\tilde \alpha}({\bf k};{\bf R})=\frac{{\tilde w^2({\bf k};{\bf R})}}{v_1({\bf R})} \ge 0
\label{I-f}
\end{equation}
is the Fourier transform of the scaled intersection
volume function (\ref{I}) and ${\tilde w}({\bf k};{\bf R})$ is the Fourier transform of the window indicator function (\ref{window}).
Again, by Parseval's theorem 
\begin{equation}
\frac{1}{(2\pi)^d} \int {\tilde \alpha}({\bf k};{\bf R}) d{\bf k}=
\frac{1}{v_1({\bf R})} \int w^2({\bf r}) d{\bf r}=1.
\end{equation}
Finally, utilizing the definition (\ref{factor}) of the structure factor,
we arrive at the Fourier representation of the number variance:
\begin{equation}
\langle N^2({\bf R}) \rangle-  \langle N({\bf R}) \rangle^2=
\langle N({\bf R}) \rangle \Bigg[\frac{1}{(2\pi)^d} \int S({\bf k}) 
{\tilde \alpha}({\bf k};{\bf R}) d{\bf k}\Bigg] 
\label{var-par}
\end{equation}
Interestingly, we  see that the variance formula can be rewritten in terms 
of the structure factor and the nonnegative function 
${\tilde \alpha}({\bf k};{\bf R})$, the Fourier transform of 
the scaled intersection volume function $\alpha({\bf r};{\bf R})$: a purely
geometric quantity.
Since the latter is independent of the correlation function $h(\bf r)$,
we conclude that the nonnegativity of the number variance
does not introduce a new realizability condition on $h(\bf r)$.
\bigskip

\noindent{\it Remarks:}
\smallskip

\noindent{1. Given the Fourier representation
formula (\ref{var-par}), it is simple to prove that the local number
variance is strictly positive for any $v_1(\bf R) > 0$.
Both the functions ${\tilde \alpha}({\bf k};{\bf R})$
 and $S(\bf k)$ are nonnegative.
Therefore, because the nonnegative integrand of formula (\ref{var-par})
cannot be zero for all $\bf k$, it immediately follows
that the local variance is strictly positive for any
statistically homogeneous point pattern whenever $v_1(\bf R) > 0$, i.e.,}
\begin{equation}
\langle N^2({\bf R}) \rangle-  \langle N({\bf R}) \rangle^2 >0.
\label{strict}
\end{equation}
\smallskip

\noindent {2. Let the window grow infinitely large in a self-similar 
(i.e., shape- and orientation-preserving) fashion. In this limit, which we will denote
simply by $v_1({\bf R}) \rightarrow \infty$, 
the function  ${\tilde \alpha}({\bf k};{\bf R})$ appearing in (\ref{var-par})
tends to $(2\pi)^d \delta({\bf k})$, where $\delta({\bf k})$ is a 
$d$-dimensional Dirac delta
function, and therefore dividing the variance (\ref{var-par}) 
by $\langle N({\bf R}) \rangle$ yields
\begin{equation}
\lim_{v_1({\bf R}) \rightarrow \infty}\frac{\displaystyle \langle N^2({\bf R}) \rangle-  \langle N({\bf R})
\rangle^2}{\displaystyle \langle N({\bf R}) \rangle}= 
S({\bf k=0})= 1+\rho\int_{\Re^d} h({\bf r}) d{\bf r}.
\label{var2}
\end{equation}
Observe also that the form of the scaled variance (\ref{var2}) for
infinitely large windows
(or infinite-wavelength limit) is identical to that for equilibrium
``open'' systems, i.e., grand canonical ensemble, in the infinite-system
limit. It is well known that the variance in the latter instance
is related to thermodynamic compressibilities or susceptibilities
\cite{Ha86}. The important
distinction is that result (\ref{var2}) is derived by considering
window fluctuations in an infinite ``closed''  possibly
{\it nonequilibrium} system.
When the point pattern comes from a statistically homogeneous equilibrium
ensemble, one can interpret the fluctuations as arising from 
differences in the point patterns in the ensemble members
for a fixed window position or, equivalently,
from moving the asymptotically large window
from point to point in a {\it single} system. The latter
scenario can be viewed as  
corresponding to density fluctuations associated with 
an ``open'' system.}

\subsection{Asymptotic Variance Formulas}

Here we apply the previous results for statistically
homogeneous point patterns to obtain asymptotic results
for large windows. The conditions under which these
expressions yield variances that only grow
as the surface area of $\Omega$ are determined.
These conditions can be expressed in terms of
spatial moments of the total correlation function $h({\bf r})$.
For simplicity, we first consider the case
of spherical windows but we show that the results
apply as well to non-spherical windows.

Many of our subsequent results will be given for a
$d$-dimensional spherical window of radius $R$ centered
at position ${\bf x}_0$. The window indicator function
becomes
\begin{equation}
w(|{\bf x}-{\bf x}_0|;R)=\Theta(R-|{\bf x}-{\bf x}_0|),
\label{window2}
\end{equation}
where $\Theta(x)$ is the Heaviside step function
\begin{equation}
\Theta(x) =\Bigg\{{0, \quad x<0,\atop{1, \quad x \ge 0.}}
\label{heaviside}
\end{equation}
Therefore, the function $v_1({\bf R})$, defined in relation (\ref{N(R)}),
becomes the volume of a spherical window 
of radius $R$ given by
\begin{equation}
v_1(R)= \frac{\pi^{d/2}}{\Gamma(1+d/2)}R^d.
\label{v1}
\end{equation}
It is convenient to introduce a dimensionless density 
$\phi$ defined by
\begin{equation}
\phi=\rho v_1(D/2)= \rho \frac{\pi^{d/2}}{2^d\Gamma(1+d/2)}D^d,
\label{phi}
\end{equation}
where $D$ is a characteristic microscopic length scale of
the system, e.g., the mean nearest-neighbor distance between 
the points.

Substitution of the expansion (\ref{alpha}) for the scaled
intersection volume $\alpha(r;R)$ into
(\ref{N2}) and assuming that the resulting integrals
separately converge, yields the variance formula for large $R$ as
\begin{equation}
\langle N^2(R) \rangle-  \langle N(R) \rangle^2=
2^d\phi\Bigg[ A\left(\frac{R}{D}\right)^d+B\left(\frac{R}{D}\right)^{d-1}+
{\ell} \left(\frac{R}{D}\right)^{d-1}\Bigg],
\label{var1}
\end{equation}
where $A$ and $B$ are the asymptotic constants given by
\begin{eqnarray}
A &=&1+\rho\int_{\Re^d} h({\bf r}) d{\bf r}=1+\frac{\phi}{v_1(D/2)} 
\int_{\Re^d} h({\bf r}) d{\bf r},
\label{A1} \\
B&=&-\frac{\phi d \Gamma(d/2)}{2Dv_1(D/2) \Gamma(\frac{d+1}{2})
\Gamma(\frac{1}{2})} \int_{\Re^d} h({\bf r})r d{\bf r},
\label{B1}
\end{eqnarray}
and ${\ell}(x)$ signifies terms of lower order than $x$ \cite{note}. In
what follows, the asymptotic constants $A$ and $B$ will generically
be referred to as ``volume'' and ``surface-area'' coefficients
for point patterns in {\it any} dimension.
\bigskip

\noindent{\it Remarks:}
\smallskip

\noindent {1. Observe that the volume coefficient $A$ is equal
to the nonnegative structure factor in the limit that the wavenumber
approaches zero, i.e.,
\begin{equation}
A=\lim_{|{\bf k}| \rightarrow 0}S({\bf k})=1+\rho\int_{\Re^d} h({\bf r}) d{\bf r} \ge 0.
\label{A3}
\end{equation}
where $S({\bf k})$ is defined by (\ref{factor}) for any dimension.
Consistent with our earlier observations about 
relation (\ref{var2}), we see that $A$ 
is the dominant term for very large windows
and indeed is the only contribution for infinitely large windows.
It is well known that point patterns generated from equilibrium
molecular systems with a wide class of interaction
potentials (e.g., hard-sphere, square-well, and
Lennard-Jones interactions) yield positive values
of $A$ in  gaseous, liquid, and many solid states.
Indeed, $A$ will be positive for
any equilibrium system possessing a positive compressibility.
This class of systems includes correlated equilibrium
particle systems, an example of which is discussed in Appendix B.
The coefficient $A$ will also be positive for a wide
class of nonequilibrium point patterns. One nonequilibrium example
is the so-called random sequential addition process \cite{To02a}.
To summarize, there is an enormously large class of point patterns
in which $A$ is nonzero.
\smallskip

\noindent {2. Because the local variance is a strictly positive
quantity for $R >0$ [cf. (\ref{strict})], we have from
(\ref{var1}) that for very large windows
\begin{equation}
A\left(\frac{R}{D}\right)^d+B\left(\frac{R}{D}\right)^{d-1} > 0.
\label{A-ineq}
\end{equation}
The crucial point to observe is that if the volume coefficient $A$
identically vanishes, then the second term within the brackets
of (\ref{var1}) dominates, and we have
the condition
\begin{equation}
B > 0,
\label{cond1}
\end{equation}
where we have used the fact that the variance  
cannot grow more slowly than $R^{d-1}$, i.e., 
the surface area of the window \cite{Be87}. 
We will refer to a system in which 
\begin{equation}
A=\lim_{|{\bf k}|\rightarrow 0}S({\bf k})=0
\end{equation}
as a ``{hyperuniform}'' system. Such point patterns do not possess
infinite-wavelength fluctuations. In a recent cosmological
study \cite{Pi02}, the term ``superhomogeneous" has
been used to describe such systems. Note that for
a one-dimensional hyperuniform system, the variance is exactly
(not asymptotically) given by
\begin{equation}
\langle N^2(R) \rangle-  \langle N(R) \rangle^2= 2\phi B,
\end{equation}
where $B$ is given by (\ref{B1}) with $d=1$, implying
that the fluctuations are bounded, i.e., do not grow with $R$ \cite{Ai01}.
\smallskip

\noindent{3. By contrast, we will refer to a point pattern 
in which the surface-area coefficient
vanishes ($B=0$) as  a ``{hyposurficial}'' system. A homogeneous
Poisson point pattern is a simple example of such a system. Inequality
(\ref{A-ineq}) in conjunction with the fact that 
the variance cannot grow more slowly than the surface area of 
a spherical (or strictly convex) 
window for statistically homogeneous and isotropic
point patterns \cite{Be87}, enables us to conclude that such a system
cannot simultaneously be hyperuniform and hyposurficial, i.e.,  the
volume coefficient $A$ [cf. (\ref{A1})] and 
surface-area coefficient $B$ [cf. (\ref{B1})] cannot
both be zero. In Appendix C,
we examine the question of how small the volume coefficient
$A$ can be made if the point pattern is hyposurficial.
\smallskip

\noindent{4. Observe also that the asymptotic variance formula (\ref{var1})
and the analysis leading to condition (\ref{cond1}) are valid for any 
statistically homogeneous
point pattern.  Now if we further assume that the point 
pattern is statistically isotropic, then the volume coefficient
(\ref{A1}) and surface-area coefficient 
(\ref{B1}) can be expressed in terms of certain moments of $h$, namely,
\begin{eqnarray}
A &=&1+ d 2^d  \phi \langle x^{d-1} \rangle,
\label{A2} \\
B&=&-\frac{ d^2 2^{d-1}\Gamma(d/2)}{ \Gamma(\frac{d+1}{2})
\Gamma(\frac{1}{2})} \phi \langle x^{d} \rangle,
\label{B2}
\end{eqnarray}
where
\begin{equation}
\langle x^n \rangle = \int_0^\infty x^n h(x) dx
\end{equation}
is the $n$th moment of the total correlation function
$h(x)$ and $x=r/D$ is a dimensionless distance.
Following the previous analysis, we see that if $A=0$,
then the condition for the variance to grow as the surface
area implies that the  the $d$th moment of $h$
must be strictly negative, i.e.,
\begin{equation}
\langle x^d \rangle < 0.
\end{equation}
}

\subsection{Direct Correlation Function and New Critical Exponents}

The {\it direct} correlation function $c(\bf r)$ of
a hyperuniform system behaves in an unconventional manner. In real space,
this function is defined by the Ornstein-Zernike equation
\begin{equation}
h({\bf r})=c({\bf r})+\rho \int_{\Re^d} h({\bf r}-{\bf r}^\prime)
c({\bf r}) d{\bf r}^\prime.
\label{OZ}
\end{equation}
This relation has primarily been used to study liquids in equilibrium
\cite{Ha86},
but it is a perfectly well-defined quantity for general (nonequilibrium)
systems, which are of central interest in this paper.
The second term is a convolution integral and therefore Fourier transforming
(\ref{OZ}) leads to
\begin{equation}
{\tilde c}({\bf k})=\frac{{\tilde h}({\bf k})}{1+\rho{\tilde h}({\bf k})},
\label{c}
\end{equation}
where ${\tilde c}({\bf k})$ is the Fourier transform of $c(\bf r)$. 
Using relation (\ref{var-par}) and definition (\ref{c}),
we can re-express the number variance for a window of arbitrary
shape in terms of the Fourier transform of the direct correlation
function as follows:
\begin{equation}
\langle N^2({\bf R}) \rangle-  \langle N({\bf R}) \rangle^2=
\langle N({\bf R}) \rangle \Bigg[\frac{1}{(2\pi)^d} \int
\frac{{\tilde \alpha}({\bf k};{\bf R})}{1-\rho {\tilde c}({\bf k})}
d{\bf k}\Bigg].
\label{var-par2}
\end{equation}

We know that for a hyperuniform system, ${\tilde h}(0)=-1/\rho$ by definition, 
i.e., the volume integral of $h(\bf r)$ exists and, in particular,
$h({\bf r})$ is a short-ranged function that decays to zero
faster than $|{\bf r}|^{-d}$. Interestingly, this
means that the denominator on the right side of
(\ref{c}) vanishes at ${\bf k=0}$ 
and therefore ${\tilde c}({\bf k=0})$ diverges to $-\infty$. 
This implies that the real-space direct correlation function
$c(\bf r)$ is long-ranged, i.e., decays slower than $|{\bf r}|^{-d}$,
and hence the volume integral of $c({\bf r})$ does not exist.
This is unconventional behavior because, in most equilibrium instances,
$c({\bf r})$ is a short-ranged function, even in the vicinity
of thermodynamic critical points where $h({\bf r})$ is long-ranged.
One can see that $c({\bf r})$
for a hyperuniform system behaves  similarly to the total correlation
function $h({\bf r})$ for an equilibrium system near
its critical point \cite{Hu87}, i.e., each of
these  functions in these respective instances become long-ranged. 
If this analogy holds, then one expects
the direct correlation function for hyperuniform
systems to have the following asymptotic
behavior for large $r\equiv |{\bf r}|$ and sufficiently large $d$:
\begin{equation}
c({\bf r}) \sim -\frac{1}{r^{d-2+\eta}} \qquad (r \rightarrow \infty),
\label{asymp2}
\end{equation}
where $(2-d) < \eta \le  2$ is a new ``critical'' exponent associated with $c(\bf r)$ for
hyperuniform systems that depends
on the space dimension \cite{footnote1}. For noninteger values of $\eta$, the asymptotic
relation (\ref{asymp2}) implies that the Fourier transform ${\tilde h}({\bf k})$
is a nonanalytic function of $k \equiv |\bf k|$. We will show in Section V
that there is a class of hyperuniform systems that obey (\ref{asymp2})
but with integer values of $\eta$, implying that ${\tilde h}({\bf k})$ is an 
analytic function of $k$. Inversion of (\ref{asymp2})
yields
\begin{equation}
{\tilde c}({\bf k}) \sim -\frac{1}{k^{2-\eta}} \qquad (k \rightarrow 0),
\label{c-eta}
\end{equation}
which, when combined with (\ref{c}), yields the asymptotic form
of the structure factor
\begin{equation}
S({\bf k}) \sim k^{2-\eta} \qquad (k \rightarrow 0).
\label{S-eta}
\end{equation}
The specific asymptotic form of $S({\bf k})$ for
small $k$ contributes to determining the ``universality'' class of
the hyperuniform system.

\begin{table}
\centering
\parbox{3.9in}{\caption{Definitions of the critical exponents
in the vicinity of or at the hyperuniform state. Here $S^{-1}(0)$
is the inverse of the structure factor at $k=0$, $\xi$ is the
correlation length, and $c(r)$ is the direct correlation function. }
\label{critical}
}
\begin{tabular}{cc}\hline\hline
Exponent& Asymptotic behavior\\ \hline 
$\gamma$ & $S^{-1}(0)  \sim  (1 - \frac{\phi}{\phi_c})^{-\gamma}$\quad ($\phi \rightarrow \phi_c^{-}$)   \\
$\gamma^{~\prime}$ & $S^{-1}(0)  \sim
(\frac{\phi}{\phi_c}-1)^{-\gamma^\prime}$\quad ($\phi \rightarrow \phi_c^{+}$)   \\
$\nu$ & $\xi  \sim  (1 - \frac{\phi}{\phi_c})^{-\nu}$\quad ($\phi \rightarrow \phi_c^{-}$)   \\
$\nu^{~\prime}$ & $\xi  \sim  (\frac{\phi}{\phi_c}-1)^{-\nu^\prime}$\quad ($\phi \rightarrow \phi_c^{+}$)   \\
$\eta$ & $c(r)  \sim  r^{2-d-\eta} $\quad ($\phi =\phi_c$)    \\ \hline\hline 
\end{tabular}
\end{table}

Let us now consider a point pattern with a reduced density
$\phi$ that is nearly hyperuniform and which can be
made hyperuniform by increasing and/or decreasing
the density. We denote by $\phi_c$ the reduced density at the hyperuniform
state. The reduced densities $\phi$ and $\phi_c$ play
the same role as temperature $T$ and critical temperature
$T_c$, respectively, in the analogous thermal problem
in the vicinity of a critical point.
Thus, we can define critical exponents associated
with the manner in which certain quantities diverge
as the critical (hyperuniform) point is approached.
For example, for $|\phi_c -\phi| \ll 1$, the inverse of the structure factor
at $k=0$, $S^{-1}(0)$, and the {\it correlation length}
$\xi$ obey the power laws
\begin{eqnarray}
S^{-1}(0) &\sim & \left(1-\frac{\phi}{\phi_c}\right)^{-\gamma}, \qquad \phi
\rightarrow \phi_c^{-}, \label{S-power}\\
\xi &\sim & \left(1-\frac{\phi}{\phi_c}\right)^{-\nu}, \qquad \phi
\rightarrow \phi_c^{-}, \label{c-power}
\end{eqnarray}
where $\gamma$ and $\nu$ are nonnegative critical exponents that
are related by the formula
\begin{equation}
\gamma=(2-\eta)\nu.
\label{inter}
\end{equation}
As will be discussed in Section V, $\xi$ characterizes the decay
of the direct correlation function in the vicinity of $\phi=\phi_c$.
Analogous critical exponents can be defined for densities
near but above $\phi_c$, as summarized in Table \ref{critical}.
In Section VB, we determine the critical exponents exactly
for certain models of disordered point patterns in $d$ dimensions.

\section{Variance Formula for a Single Point Pattern}

In this section, we derive a new formula for the number variance 
of a single realization of a point pattern consisting of a large
number of points $N$  
in a large system of volume $V$.
This is necessarily a
volume-average formulation.
Fluctuations for a fixed window size arise because
we let the window uniformly sample the space. As we will
show, depending on the nature of the point pattern,
this formula will generally lead to a result that
is different from formula (\ref{N2}), which
was derived for a statistically homogeneous system.
We also show that the formula derived here is
preferable for finding point patterns  with
an extremal or targeted value of the number variance.

For notational simplicity,
we consider a $d$-dimensional spherical window of radius
$R$, keeping in mind that the results of this section
apply as well (with obvious notational changes) to regular domains 
of arbitrary shape. 
We assume that the characteristic size of the system 
is much larger than the window radius 
so that boundary effects can be
neglected and that the large numbers $N \gg 1$ and $V\gg 1$ are
comparable such that
$\rho\equiv N/V$ is  a finite number density.
Let us recall relation (\ref{N}) for the number of
points $N({\bf x}_0;R)$ contained within a window 
at position ${\bf x}_0$ in a system of
volume $V$ in which there are $N$ points.    
We let the window uniformly sample
the space and define the average number of
points within the window to be
\begin{eqnarray}
{\overline {N(R)}} &\equiv& \frac{1}{V} \int_V \sum_{i=1}^N w(|{\bf r}_i -{\bf x_0}|;R) d{\bf x}_0 \nonumber \\
&=& \rho \int_V  \Theta(R-r) d{\bf r} \nonumber \\
&=& \rho v_1(R) \nonumber \\
&=& 2^d \phi \left(\frac{R}{D}\right)^d,
\label{N-avg}
\end{eqnarray}
where $v_1(r)$ and $\phi$ are given by (\ref{v1})
and  (\ref{phi}), respectively.

Similarly, squaring relation (\ref{N}) and averaging yields
\begin{eqnarray}
{\overline {N^2(R)}} &=& \frac{1}{V} \int_V \sum_{i=1}^N w(|{\bf r}_i -{\bf
x_0}|;R) d{\bf x}_0 + \frac{1}{V} \int_V \sum_{i\neq j}^N 
w(|{\bf r}_i -{\bf x_0}|;R) w(|{\bf r}_j -{\bf x_0}|;R) d{\bf x}_0 \nonumber \\
&=& \rho v_1(R)+ \frac{\rho v_1(R)}{N} \sum_{i \neq j}^N \alpha(r_{ij};R),
\end{eqnarray}
where $\alpha(r;R)$ is the scaled intersection volume, given explicitly 
by (\ref{I2}), and $r_{ij}= |{\bf r}_i -{\bf r}_j|$.
Therefore, the local variance $\sigma^2(R)$ is  given by
\begin{eqnarray}
\sigma^2(R)\equiv {\overline {N^2(R)}}-{\overline {N(R)}}^2 
&=&  {\overline {N(R)}} \Bigg[1 -\rho v_1(R) +
\frac{1}{N}\sum_{i \neq j}^N \alpha(r_{ij};R) \Bigg] \nonumber \\
&=&2^d \phi \left(\frac{R}{D}\right)^d \Bigg[ 1 -2^d \phi \left(\frac{R}{D}\right)^d + 
\frac{1}{N}\sum_{i \neq j}^N \alpha(r_{ij};R) \Bigg].
\label{vol-var}
\end{eqnarray}
The last term within the brackets
is the sum of scaled intersection volumes between all 
point pairs, per point.   
\bigskip

\noindent{\it Remarks:}
\smallskip

\noindent{1. It is important
to observe that the series in (\ref{vol-var}) terminates
for $r_{ij} > 2R$ even for infinitely large systems.}
\smallskip

\noindent {2. Note that the variance formula (\ref{vol-var}) is different from
the ensemble-average formula (\ref{N2}), which involves an additional
weighted average over pairs of points; thus, the appearance
of the total correlation function $h(\bf r)$. Therefore,
the variance function (\ref{vol-var}), unlike the
variance function (\ref{N2}), will generally contain small-scale fluctuations 
with respect to $R$, of wavelength on the order
of the mean separation between the points, that are superposed on the
large-scale variations with respect to $R$ (see examples in Section IV). 
The expressions (\ref{vol-var})  and (\ref{N2}) are
identically the same for statistically homogeneous (infinite) systems, in which
case the amplitudes of the small-scale fluctuations vanish.
}
\smallskip

\noindent {3. Because the variance formula is valid for a single
realization, one can use it, in principle, to find the particular
point pattern which minimizes the variance at a fixed value
of $R$. In other words,  it is desired
to minimize $\sigma^2(R)$ for a particular value of $R$
among all $r_{ij} \le 2R$, i.e.,
\begin{equation}
\min_{\forall r_{ij} \le 2R} \sigma^2(R)
\end{equation}
where $\sigma^2(R)$ is given by (\ref{vol-var}). The scaled
intersection volume $\alpha(r_{ij};R)$ appearing in (\ref{vol-var})
is a nonnegative function of $r_{ij}$ (see Fig. \ref{intersection}) and 
can be viewed as
a {\it repulsive} pair potential between a point $i$ and a point $j$.
Finding the global minimum of $\sigma^2(R)$ is equivalent
to determining the ground state for the ``potential energy" function
represented by the pairwise sum in (\ref{vol-var}).
Such global optimization problems can be attacked using
simulated annealing techniques, for example.
More generally, one could devise an optimization
scheme in which a {\it targeted} value of the variance (rather than
an extremal value) is sought \cite{To}.

\smallskip

\noindent{4. Because the pairwise sum in (\ref{vol-var}) is positive, 
we immediately obtain from (\ref{vol-var}) the following lower
bound on the variance:
\begin{equation}
\sigma^2(R) \ge 
2^d \phi \left(\frac{R}{D}\right)^d \Bigg[ 1 -2^d \phi
\left(\frac{R}{D}\right)^d \Bigg].
\end{equation}
This bound is exact for $R \le r_{\mbox{min}}/2$, where 
$r_{\mbox{min}}$ is the minimum pairwise distance, and therefore
provides an accurate estimate of the variance for small $R$.
For sufficiently large $R$, however, the bound becomes negative
and therefore provides a poor estimate of the variance.
\smallskip

\noindent {5. For large $R$ in the special case of hyperuniform systems, 
the large-scale variations in $R$ will grow as $R^{d-1}$, and
so we have from (\ref{vol-var}) that
\begin{equation}
\sigma^2(R) = \Lambda(R) \left(\frac{R}{D}\right)^{d-1}+ 
{\cal O}\left(\frac{R}{D}\right)^{d-2}
\label{sig-asymp}
\end{equation}
where 
\begin{equation}
\Lambda(R)= 2^d \phi \left(\frac{R}{D}\right) \Bigg[ 1-2^d \phi \left(\frac{R}{D}\right)^d + 
\frac{1}{N}\sum_{i \neq j}^N \alpha(r_{ij};R) \Bigg]
\label{Lambda(R)}
\end{equation}
is the asymptotic ``surface-area''  function that contains the small-scale variations in $R$.}
\smallskip

\noindent{6. It is useful to average the small-scale 
function $\Lambda(R)$ over $R$ to yield
the constant
\begin{equation}
{\overline \Lambda}(L)= \frac{1}{L} \int_0^L \Lambda(R) dR,
\label{S}
\end{equation} 
where $\Lambda(R)$ is given by (\ref{Lambda(R)}). In the case of a statistically
homogeneous system,  the constant surface-area coefficient
\begin{equation}
{\overline \Lambda}\equiv \lim_{L \rightarrow \infty} {\overline \Lambda}(L)=\lim_{L \rightarrow \infty}\frac{1}{L} \int_0^L \Lambda(R) dR
\label{S-constant}
\end{equation}
is trivially related to the surface-area coefficient $B$, 
defined by (\ref{B1}) in the asymptotic ensemble-average
formula, by the expression
\begin{equation}
{\overline \Lambda} = 2^d \phi B=\frac{-~2^{d-1}\phi^2 d \Gamma(d/2)}{Dv_1(D/2)
\Gamma(\frac{d+1}{2})
\Gamma(\frac{1}{2})} \int_{\Re^d} h({\bf r})r d{\bf r}.
\label{S-ensemble}
\end{equation}
\smallskip

\noindent {7. Because the formula for the
coefficient ${\overline \Lambda}$ is defined 
for a single
realization, we can employ it to obtain the particular
point pattern that minimizes it. Thus, the optimization
problem is the following:
\begin{equation}
\min_{\mbox{\scriptsize all } r_{ij} \le 2L} {\overline \Lambda},
\end{equation}
where ${\overline \Lambda}$ is given by (\ref{S}).
\smallskip

\noindent {8. For large systems in which any point ``sees'' 
an environment typical of all points, relation (\ref{vol-var}) for the variance
can be simplified. This requirement is met by all
infinite periodic lattices for any $R$ as well as statistically homogeneous
point patterns for sufficiently large $R$. In such instances, the second term within the brackets
of (\ref{vol-var})  can be written as sum of scaled intersection
volumes over $N-1$ points and some reference point.
Thus, we can rewrite the variance as
\begin{equation}
\sigma^2(R)= 2^d \phi \left(\frac{R}{D}\right)^d \Bigg[ 1 -2^d \phi
\left(\frac{R}{D}\right)^d  +
\sum_{k=1}^{N-1} \alpha(r_k;R)\Bigg],
\label{vol-var2}
\end{equation}
where $r_k$ is the distance from the reference point 
to the $k$th point.  The asymptotic expression  (\ref{sig-asymp}) for $\sigma^2(R)$
and relation (\ref{S}) for ${\overline \Lambda}(R)$ 
still apply but with $\Lambda(R)$ given by the simpler formula
\begin{equation}
\Lambda(R)= 2^d \phi \left(\frac{R}{D}\right) \Bigg[ 1 -2^d \phi \left(\frac{R}{D}\right)^d + 
\sum_{k=1}^{N-1} \alpha(r_k;R) \Bigg]
\label{Lambda(R)-2}
\end{equation}
We emphasize that the simplified formulas (\ref{vol-var2}) and
(\ref{Lambda(R)-2}) {\it cannot} be used for the aforementioned 
optimization calculations. The latter requires the full
pairwise sum appearing in the general relation (\ref{vol-var}).
\smallskip

\noindent 9. In order to make the surface-area function
$\Lambda(R)$ or surface-area coefficient ${\overline \Lambda}$ independent of
the characteristic length scale or, equivalently, density of the
hyperuniform point pattern, one can divide 
each of these quantities by $\phi^{(d-1)/d}$, i.e.,
\begin{equation}
\frac{\Lambda(R)}{\phi^{(d-1)/d}}, \qquad \mbox{or} \qquad 
\frac{\overline \Lambda}{\phi^{(d-1)/d}}.
\label{Lambda-scaling}
\end{equation}
This scaling arises by recognizing that normalization
of the asymptotic relation (\ref{sig-asymp}) by expression (\ref{N-avg}) for
$\langle {\overline N(R)} \rangle$ taken to the power
$(d-1)/d$ renders the resulting normalized relation independent
of $R/D$. Such a scaling will be used to compare
calculations of $\Lambda(R)$ and ${\overline \Lambda}$ 
for different ordered and disordered point patterns
to one another in the subsequent sections.
Note that since one-dimensional hyperuniform patterns
have bounded fluctuations, this scaling is irrelevant
for $d=1$.

\section{Calculations for Infinite Periodic Lattices}

It is useful and instructive to compute the variance, using the formulas
derived in the previous section, for common infinite periodic
lattices, which are hyperuniform systems. To our knowledge, 
explicit calculations have only been obtained
for the square lattice \cite{Ken48}
and triangular lattice \cite{Ken53} in two dimensions. 
Here we will obtain explicit results for other two-dimensional
lattices as well as one- and three-dimensional lattices.
We take the window to be a $d$-dimensional sphere of radius $R$.

For infinite periodic lattices, Fourier analysis leads to an alternative
representation of the variance. Let the sites of the lattice
be specified by the primitive lattice vector ${\bf P}$ defined
by the expression
\begin{equation}
{\bf P}= n_1 {\bf a}_1+ n_2 {\bf a}_2+ \cdots + n_{d-1} {\bf a}_{d-1}+n_d {\bf a}_d,
\end{equation}
where ${\bf a}_i$ are the basis vectors of the unit cell array
and $n_i$ spans all the integers for $i=1,2,\cdots d$. Denote by $U$ the unit
cell and $v_C$ its volume. It is clear that  the number of points $N({\bf x}_0; R)$
 within the window at ${\bf x}_0$ [cf. (\ref{N})] in this instance becomes
\begin{equation}
N({\bf x}_0; R)= \sum_{\bf P} \Theta(R-|{\bf P}-{\bf x}_0|),
\end{equation}
where the sum is over all $\bf P$.

The number $N({\bf x}_0; R)$ is a periodic function in the window position
${\bf x}_0$ and therefore it can be expanded in a Fourier series as
\begin{equation}
N({\bf x}_0; R)=\rho v_1(R)+ \sum_{\bf q \neq 0} a({\bf q})
e^{\displaystyle i {\bf q} 
\cdot {\bf x}_0}
\label{N-per}
\end{equation}
where $\bf q$ is the reciprocal lattice vector such that ${\bf q}\cdot {\bf P}=2\pi m$
(where $m= \pm 1, \pm 2, \pm 3 \cdots$) and the sum 
is over all $\bf q$ except $\bf q=0$. Following Kendall and Rankin \cite{Ken53},
the coefficients $a(\bf q)$, for $\bf q \neq 0$, are given by 
\begin{eqnarray}
a({\bf q})&=& \frac{1}{v_C} \int_{U} N({\bf x}_0; R)
e^{\displaystyle -i {\bf q}  \cdot {\bf x}_0} d{\bf x}_0 \nonumber\\
&=& \frac{1}{v_C} \sum_{\bf P}\int_U
  \Theta(R-|{\bf P}-{\bf x}_0|)
e^{\displaystyle -i {\bf q}  \cdot {\bf x}_0} d{\bf x}_0 \nonumber\\
&=&\frac{1}{v_C} \int_{\Re^d} \Theta(R-|{\bf T}|) e^{\displaystyle i {\bf q}  \cdot {\bf T}} d{\bf T}\nonumber\\
&=& \frac{1}{v_C}\left(\frac{2\pi}{qR}\right)^{d/2} R^d J_{d/2}(qR),
\label{a}
\end{eqnarray}
where $J_{\nu}(x)$ is the Bessel function of order $\nu$.
Note that the integral in the third line is nothing more than
the Fourier transform of the window indicator function, which is given
by (\ref{window-Fourier}) in Appendix A. The analysis above assumes
that there is one point per unit cell, i.e., we have considered
Bravais lattices. One can easily generalize 
it to the case of an arbitrary number of points $n_C$ per unit cell.
Formula (\ref{a}) would then involve  $n_C-1$  additional terms
of similar form to the original one.

By Parseval's theorem for Fourier series, the number variance
$\sigma^2(R)$ is given explicitly by
\begin{eqnarray}
\sigma^2(R) &\equiv& \frac{1}{v_C} \int_{U} [N({\bf x}_0;R)-\rho v_1(R)]^2
d{\bf x}_0 \nonumber \\
 &=& \sum_{\bf q \neq 0} a^2({\bf q}) \nonumber \\
&=& \frac{R^d}{v_C^2} \sum_{\bf q \neq 0}
\left(\frac{2\pi}{q}\right)^{d}  [J_{d/2}(qR)]^2.
\label{parseval}
\end{eqnarray}
One can easily obtain
an asymptotic expression for the variance for large
$R$ by replacing the Bessel function in (\ref{parseval}) by the first
term of its asymptotic expansion, and thus we have
\begin{equation}
\sigma^2(R) = \Lambda(R) \left(\frac{R}{D}\right)^{d-1}+ 
{\cal O}\left(\frac{R}{D}\right)^{d-2}
\end{equation}
where $D$ is a characteristic microscopic length scale, say the lattice spacing, and 
\begin{equation}
\Lambda(R)=\frac{2^{d+1} \pi^{d-1}D^{2d}}{v_C^2} \displaystyle\sum_{\bf q \neq 0} 
\frac{\cos^2\Big[qR -(d+1)\pi/4\Big]}{(qD)^{d+1}},
\label{S-period}
\end{equation}
describes small-scale variations in $R$. As before, it is convenient to
compute the average of $\Lambda(R)$ over $R$ to give the surface-area coefficient:
\begin{eqnarray}
{\overline \Lambda} &=& \lim_{L \rightarrow \infty}\frac{1}{L} \int_0^L \Lambda(R) dR \nonumber \\
&=& \frac{2^{d} \pi^{d-1}D^{2d}}{v_C^2}  \sum_{\bf q \neq 0} \frac{1}{(qD)^{d+1}}.
\label{S-avg}
\end{eqnarray}

It is useful here to apply the specialized volume-average formula
(\ref{vol-var2}) to the case of infinite periodic lattices. Recognizing
that the configuration of an infinite periodic point pattern may
be characterized by the distances $r_k$ 
and coordination numbers $Z_k$ for the successive shells of neighbors
($k=1,2,3,\ldots$) from a lattice point, we find from
(\ref{vol-var2}) that the variance 
can also be represented as
\begin{equation}
\sigma^2(R)= 2^d \phi \left(\frac{R}{D}\right)^d \Bigg[ 1 -2^d \phi
\left(\frac{R}{D}\right)^d +
\sum_{k=1}^{\infty} Z_k\alpha(r_k;R) \Bigg].
\label{vol-var3}
\end{equation}
The asymptotic expression  (\ref{sig-asymp}) for $\sigma^2(R)$
and relation (\ref{S}) for the surface-area coefficient ${\overline \Lambda}(R)$
still apply but with $\Lambda(R)$ given by 
\begin{equation}
\Lambda(R)= 2^d \phi \left(\frac{R}{D}\right) \Bigg[ 1 -2^d \phi \left(\frac{R}{D}\right)^d +
\sum_{k=1}^{\infty} Z_k\alpha(r_k;R)
\Bigg].
\label{Lambda(R)-3}
\end{equation}
Formula (\ref{vol-var3}) was obtained by Kendall and Rankin \cite{Ken53}
using a more complicated derivation. Moreover, their derivation
only applies to periodic point patterns. Our more general
formula (\ref{vol-var2}) is also valid for statistically
homogeneous point patterns. We also note that our most general volume-average
representation (\ref{vol-var}) of the variance, from which
formula (\ref{vol-var2}) is derived, is applicable to 
arbitrary point patterns and its derivation is quite straightforward.

One can also evaluate the asymptotic coefficient ${\overline \Lambda}$
using the ensemble-average formula (\ref{S-ensemble}). Strictly speaking,
this formula is not applicable to periodic point patterns
because such systems are not statistically homogeneous (neither
are they statistically isotropic). To see
the potential problem that arises by naively applying
(\ref{S-ensemble}), let the origin be a lattice point
in the system and consider determining the radial distribution function
$g_2(r)$ by counting the number of lattice points at a radial distance
$r_k$ from the origin. For a lattice in $d$ dimensions,
we have that
\begin{equation}
g_2(r)=\sum_{k=1}^\infty \frac{Z_k \delta(r-r_k)}{\rho s_1(r_k)},
\end{equation}
where $s_1(r)$ is the surface area of a sphere of radius $r$
given by (\ref{area-sph}) and $Z_k$ is
the coordination number of the $k$th shell. 
It is seen that substitution of the corresponding
total correlation function $h(r)\equiv g_2(r)-1$ into
(\ref{S-ensemble}) results in a nonconvergent sum. However, using
a convergence ``trick'' \cite{St98}, one can properly
assure a convergent expression by reinterpreting the surface-area coefficient
(\ref{S-ensemble}) for a periodic lattice in the following manner:
\begin{eqnarray}
{\overline \Lambda} &=&\lim_{ \beta\rightarrow 0^+} 
\frac{- ~2^{d-1}\phi^2 d \Gamma(d/2)}{Dv_1(D/2)
\Gamma(\frac{d+1}{2})
\Gamma(\frac{1}{2})} \int_{\Re^d} e^{\displaystyle -\beta r^2}h(r)r d{\bf r} \nonumber \\
&=& \lim_{ \beta\rightarrow 0^+} 
\frac{2^{d-1}\phi d}{D 
\Gamma(\frac{1}{2})}\left[  \frac{\phi \pi^{d/2}}{v_1(D/2) 
\beta^{\frac{d+1}{2}}}- \frac{\Gamma(d/2)}{\Gamma(\frac{d+1}{2})}
\sum_{k=1}^{\infty} Z_k r_k e^{\displaystyle-\beta
r_k^2}\right].
\label{S-ensemble2}
\end{eqnarray}

\subsection{One-Dimensional Examples}


Here we obtain exact expressions for the number of points
and number variance for general one-dimensional periodic point
patterns using the aforementioned Fourier analysis. 
Using this result, we prove that the simple periodic linear array
corresponds to the global minimum in $\overline \Lambda$.
Subsequently, we employ the volume-average and ensemble-average
formulations of Sections II and III to obtain some of the same results
in order to compare the three different methods.
Recall that hyperuniform systems in one dimension have
bounded fluctuations.

\begin{figure}[bthp]
\centerline{\psfig{file=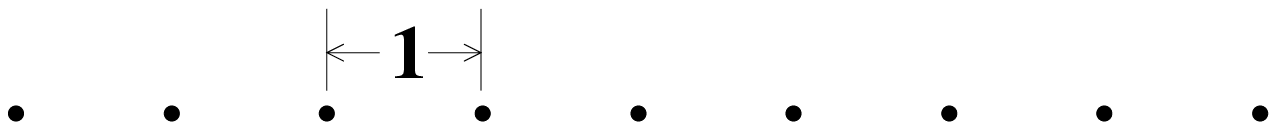,width=3.0in}}
\vspace{0.2in}
\centerline{\psfig{file=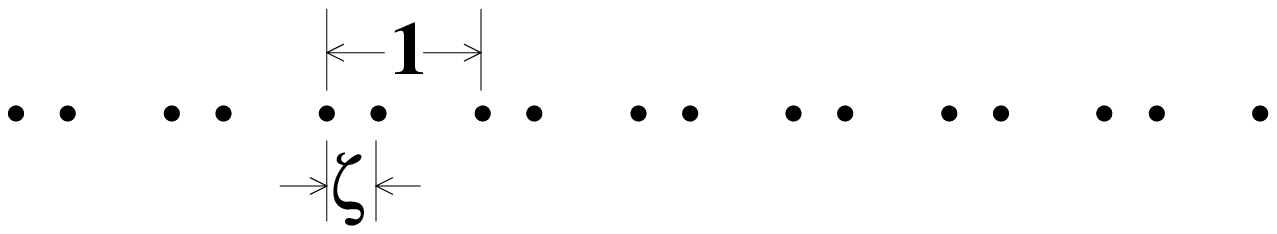,width=3.0in}}
\caption{Portions of two  one-dimensional periodic point patterns, where $v_C=D=1$.
The top and bottom arrays are the single-scale and  two-scale examples, respectively.} 
\label{1-scale}
\end{figure}

\begin{figure}[bthp]
\centerline{\psfig{file=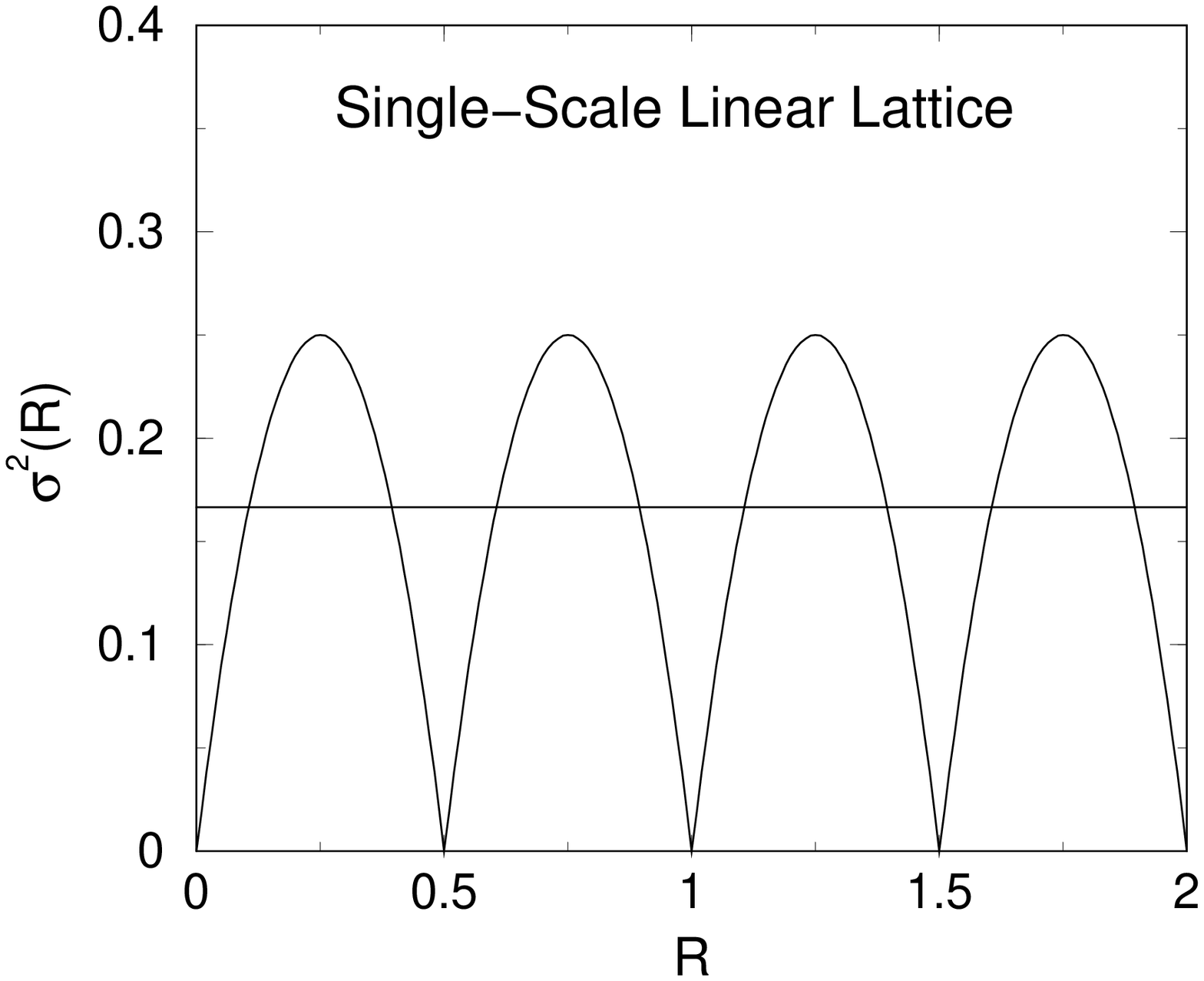,width=3.0in}\hspace{0.3in}\psfig{file=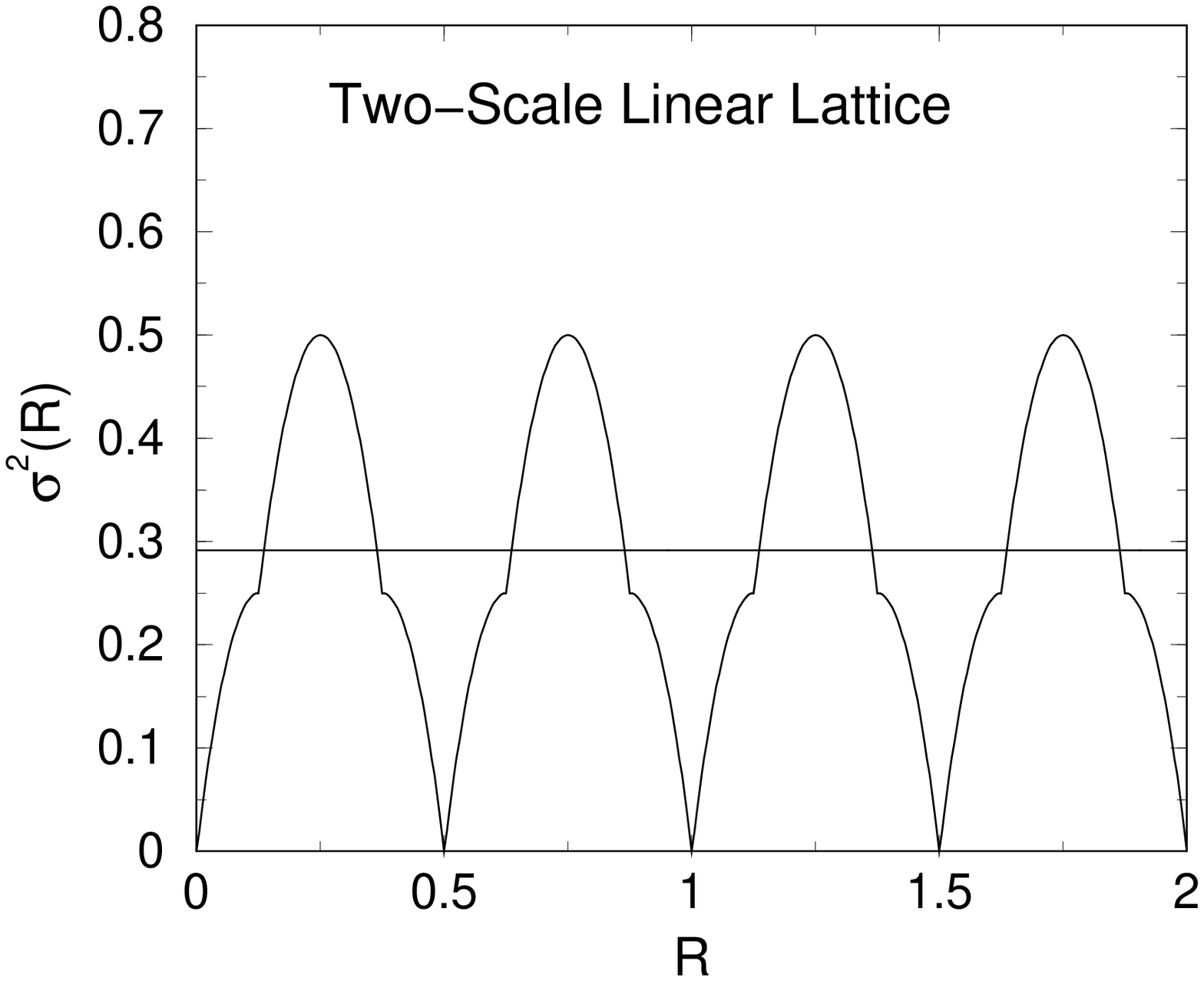,width=3.0in}}
\vspace{-0.1in}
\caption{Left panel: 
The quadratic periodic variance function $\sigma^2(R)$  for the
single-scale periodic one-dimensional point pattern given by (\ref{var-single}). The horizontal
line is the average ${\overline \Lambda}=1/6$. Right panel:
The piecewise-quadratic periodic variance function $\sigma^2(R)$ for the two-scale periodic one-dimensional point 
pattern given by (\ref{var-2}) for the case $\zeta=1/4$. The horizontal
line is the average ${\overline \Lambda}=7/24$.} 
\label{single-scale}
\end{figure}

Let us first consider the simplest periodic point pattern
in which each point is equi-distant from its 
near neighbors (see Fig. \ref{1-scale})
and let this nearest-neighbor distance be unity ($v_C=D=1$). Applying
relations (\ref{N-per}) and (\ref{a}) and recognizing that 
${\bf q}= 2\pi m {\bf a}_1/D$ ($m= \pm 1, \pm 2, \cdots$) 
for nonzero $\bf q$ yields that the number of points
contained within a one-dimensional window of radius $R$
centered at position $x_0$:
\begin{equation}
N(R;x_0)=2R+ \frac{2}{\pi} \sum_{m=1}^\infty \frac{\sin (2\pi m R)
\cos(2\pi m x_0)}{m}.
\end{equation}
According to relation (\ref{parseval}), the associated variance is given by
\begin{equation}
\sigma^2(R)=\frac{2}{\pi^2} \sum_{m=1}^\infty \frac{\sin^2(2\pi m R)}{m^2}.
\label{var-single}
\end{equation}
The variance $\sigma^2(R)$ is a periodic function with
period $1/2$ and is equal to the quadratic function
$2R(1-2R)$ for $0\le R \le 1/2$ (see Fig.
\ref{single-scale}).
Finally, the surface-area coefficient ${\overline \Lambda}$, defined by (\ref{S-avg}), 
which in one dimension amounts
to the positional average of the variance for any value of $R$, is
exactly given by
the constant
\begin{equation}
{\overline \Lambda}= \frac{1}{\pi^2} \sum_{m=1}^\infty \frac{1}{m^2}=\frac{1}{6}.
\label{1/6}
\end{equation}

It is known that this simple linear array
yields the minimum value of ${\overline \Lambda}$ among all
one-dimensional regular lattices. This is intuitively clear 
from the volume-average variance relation (\ref{vol-var})
for $d=1$; the  linear repulsive effective ``pair potential'' contained 
therein is evidently responsible for such a minimum.
However, heretofore it was not known whether this pattern corresponded to a global 
minimum, i.e., the smallest value of ${\overline \Lambda}$ among all infinite one-dimensional
hyperuniform patterns. We now prove that the single-scale lattice
indeed produces the global minimum. To prove this assertion,
we utilize the identity
\begin{equation}
f(x)=\frac{1}{\pi^2}\sum_{m=1}^{\infty}\frac{1+2 \cos (2\pi m x)}{m^2}=
\frac{1}{2}-2x(1-x),
\label{convex}
\end{equation}
and note that $f(x)$ is a convex quadratic nonnegative function
for all real $x$. Now consider a case in which there are $M$  points per 
unit cell in which the length of the unit cell is still
unity. Thus, excluding the point at each lattice site, there are $M-1$ 
points inside the unit cell with positions $\zeta_1,\zeta_2,\ldots,\zeta_{M-1}$  
such that each $\zeta_i$ lies in the interval $(0,1)$. Without loss of 
generality, we arrange the $M-1$ points such that $\zeta_i < \zeta_{i+1}$ 
($i=1,2,\ldots,M-2$), {\it but their
positions are otherwise arbitrary}. 
Following a similar analysis as the one above, we find that 
the number of points within a window centered at $x_0$ is exactly given by
\begin{equation}
N(R;x_0)=2M R+ 2 \sum_{m=1}^\infty \frac{ \sin(2\pi m R)\Big[\sum_{j=0}^{M-1}
\cos [2\pi m (x_0-\zeta_j)]\Big]}{m}
\end{equation}
where $\zeta_0 \equiv 0$. The variance is therefore given by
\begin{equation}
\sigma^2(R)= \frac{2}{\pi^2} \sum_{m=1}^\infty 
\frac{\sin^2(2\pi m R)\Big[M+ \sum_{j=1}^{M-1}\cos (2\pi m \zeta_j)
+\sum_{j<k}^{M-1}\cos [2\pi m (\zeta_k-\zeta_j)]  \Big]}{m^2}.
\label{var-2}
\end{equation}
We see that the variance $\sigma^2(R)$ for an arbitrary one-dimensional point pattern
within the unit cell is a periodic function with period 
$1/2$. (As we will see,  the variance in higher dimensions
is not a periodic function in $R$ for periodic point patterns.)
The average of the variance is exactly equal
to the surface-area coefficient (\ref{S-avg}):
\begin{eqnarray}
{\overline \Lambda} &=& \frac{1}{\pi^2} \sum_{m=1}^\infty 
\frac{ M+ \sum_{j=1}^{M-1}\cos (2\pi m \zeta_j)
+\sum_{j<k}^{M-1}\cos [2\pi m (\zeta_k-\zeta_j)]  }{m^2} \nonumber \\
&=& -\frac{M(M-3)}{12}+\sum_{j=1}^{M-1} f(\zeta_j)
+\sum_{j<k}^{M-1} f(\zeta_k-\zeta_j),
\label{var-2b}
\end{eqnarray}
where $f(x)$ is given by (\ref{convex}). Because ${\overline \Lambda}$
is given by a sum of convex quadratic nonnegative functions, the global
minimum is found from the zeroes of the derivative $\partial {\overline
\Lambda}/\partial \zeta_n$:
\begin{equation}
\frac{\partial {\overline \Lambda}}{\partial \zeta_n}=
0=1-2\zeta_n +\sum_{j=1}^{n-1} (1-2\zeta_n+2\zeta_j)
-\sum_{j=n+1}^{M-1} (1+2\zeta_n-2\zeta_j), \qquad (n=1,2,\ldots,M-1).
\label{diff}
\end{equation}
It is easy to verify that the global minimum 
is achieved when 
the $M-1$ are uniformly distributed in the interval
($0,1$), i.e., $\zeta_n=n/M$ ($n=1,2,\ldots,M-1$), yielding
$\overline \Lambda=1/6$.
Since this result is valid for arbitrary $M$, the simple
single-scale lattice produces the global minimum value
of ${\overline \Lambda}$ among all infinite one-dimensional hyperuniform
point patterns.

Note that the single-scale
lattice corresponds to the densest packing of
one-dimensional congruent hard spheres (rods) on the real
line. This might lead one to conjecture
that the Bravais lattice associated with the densest
packing of congruent spheres in
any space dimension $d$ provides the minimal value 
of $\overline \Lambda$ among all periodic
lattices for spherical windows. As we will see, this
turns out to be the case for $d=2$, but not for
$d=3$.

\begin{table}[bthp]
\caption{ The surface-area coefficient ${\overline \Lambda}$
for some ordered and disordered one-dimensional point patterns.
The result for the two-scale lattice is for $\zeta \equiv \zeta_1=0.25$.}
\vspace{0.05in}
\begin{tabular}{c|c|c} \hline \hline 
Pattern  & $\phi$ & ${\overline \Lambda}$  
\\ \hline \hline
Single-Scale Lattice & 1 &$1/6 \approx 0.166667$\\ \hline 
Step+Delta-Function $g_2$  & 0.75& $3/16=0.1875$\\ \hline 
Step-Function $g_2$& 0.5 & $1/4=0.25$ \\ \hline
Two-Scale Lattice & 2 & $7/24 \approx 0.291667$\\ \hline
Lattice-Gas & 1& $1/3 \approx 0.333333$ \\ \hline \hline
\end{tabular}
\label{1d}
\end{table}

The variance as computed from (\ref{var-2}) for the case $M=2$, 
which we call the ``two-scale'' lattice (see Fig. \ref{1-scale}), is included
in Fig. \ref{single-scale} for $\zeta\equiv \zeta_1=1/4$.
In this instance, ${\overline \Lambda}=7/24$. Clearly, the variance for the two-scale
case bounds from above the variance for the single-scale case. 
Table \ref{1d} compares the surface-area coefficient for the single-scale and two-scale
one-dimensional lattices. The other one-dimensional results
summarized in Table \ref{1d} will be discussed in 
the ensuing sections.  The potential use of the local
variance as an order metric for hyperuniform point patterns
in any dimension is discussed in Section VI.

Consider obtaining the volume-average representation of the
variance for the two aforementioned one-dimensional periodic
patterns from (\ref{vol-var3}). Using relations (\ref{I2}) and
(\ref{inter-v2-1d}) for the scaled intersection volume $\alpha(r;R)$, 
we find for any one-dimensional
periodic point pattern in which $D=1$ that
\begin{equation}
\sigma^2(R)= 2\phi R \Bigg[ 1-2 \phi R+
\sum_{k=1}^{M_R} Z_k\left(1-\frac{r_k}{2R}\right)\Theta(2R-r_k)
 \Bigg],
\label{vol-rods1}
\end{equation}
where $M_R$ corresponds to the largest value of $k$ for which $r_k <2R$.
Because in one dimension $\Lambda(R)=\sigma^2(R)$, where $\Lambda(R)$ is the 
function defined by (\ref{Lambda(R)-3}), it follows that the
average ${\overline \Lambda}$ is given by
\begin{eqnarray}
{\overline \Lambda} &=&  2 \int_0^{1/2} \Lambda(R) dR =
 \phi L \Bigg[\left( 1- \frac{4\phi L}{3}\right)+ \sum_{k=1}^{M_L}
Z_k\left(1-\frac{r_k}{2L}\right)^2\Bigg],
\label{vol-rods2}
\end{eqnarray}
where $M_L$ corresponds to the largest value of $k$ for which $r_k <2L$.
Using the fact that $\phi=1$, $r_k=k$, and $Z_k=2$ for all $k$ for
the single-scale lattice, one can
easily reproduce the graph for $\sigma^2(R)$ depicted in 
Fig. \ref{single-scale} using relation (\ref{vol-rods1}) 
and verify that  ${\overline \Lambda}=1/6$ employing
relation (\ref{vol-rods2}). Similarly,
for the two-scale case, we have that 
$\phi=2$, $r_k=k/4$ and $Z_k=1$ for odd $k$, and $r_k=k/2$ 
and $Z_k=2$ for even  $k$. Hence, relation (\ref{vol-rods1}) 
leads to the same graph of the variance shown in Fig. \ref{single-scale}, 
and relation (\ref{vol-rods2})
yields ${\overline \Lambda}=7/24$ for $\zeta=0.25$, as before.

We can also compute the surface-area coefficient using the
ensemble-average relation (\ref{S-ensemble2}).
In one dimension, this relation yields
\begin{equation}
{\overline \Lambda} = \lim_{ \beta\rightarrow 0^+} \left[\frac{\phi^2}{\beta}- 
\phi\sum_{k=1}^{\infty} Z_k r_k e^{\displaystyle-\beta r_k^2}  \right],
\label{sum}
\end{equation}
where we have taken $D=1$. 
The sum in (\ref{sum}) can be evaluated exactly using 
 the Euler-Maclaurin summation formula \cite{Ab72}.
If $f(k)$ is a function defined on the integers,
and continuous and differentiable in between, the Euler-Maclaurin summation
formula provides an asymptotic expansion of the sum
$\sum_{k=0}^n f(k)$ as $n \rightarrow \infty$.
Applying this asymptotic formula to (\ref{sum}) in the cases of the single-scale and two-scale lattices,
yields that ${\overline \Lambda}=1/6$ and ${\overline \Lambda}=7/24$, respectively,
which agree with the results obtained using the previous two methods.
Although the Fourier-analysis and volume-average procedures
are more direct methods to determine ${\overline \Lambda}$ for one-dimensional lattices, we will
see that the representation (\ref{S-ensemble2}) provides an efficient means
of computing  ${\overline \Lambda}$ for lattices in higher dimensions.

\begin{figure}[bthp]
\centerline{\psfig{file=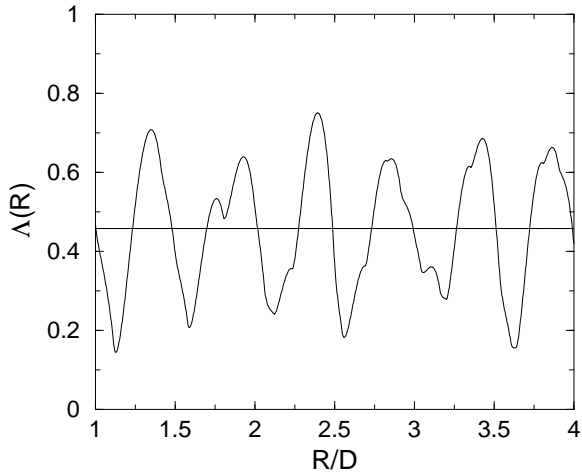,width=3.0in}}
\vspace{-0.2in}
\caption{The asymptotic surface-area function $\Lambda(R)$ for the square lattice
for $1 \le R \le 4$, where  $D$ is the lattice spacing. The horizontal
line is the asymptotic average value ${\overline \Lambda}=0.457649$.} 
\label{square2}
\end{figure}

\subsection{Two-Dimensional Examples}
\vspace{-0.2in}

Here we evaluate variance characteristics
for the following four common two-dimensional
lattices: square, triangular, honeycomb, and Kagom{\' e} lattices.
From  the lattice series (\ref{parseval}),
(\ref{S-period}) and (\ref{S-avg}) with $d=2$, we have general two-dimensional
series relations for the variance $\sigma^2(R)$, asymptotic surface-area
function $\Lambda(R)$ and surface-area coefficient
${\overline \Lambda}$, respectively.
For a specific lattice, the evaluation of any of these series requires
the reciprocal lattice vector $\bf q$ and $v_C$. 
For example, for the square lattice,
${\bf q}=2\pi [m_1 {\bf a}_1
+m_2 {\bf a}_2]/D$
($m_i= 0,\pm 1, \pm 2, \cdots$) for nonzero $\bf q$ and $v_C=D^2$. The sums
are straightforward to evaluate, even if they converge slowly.
Provided that $R$ is not very large, however, the 
corresponding volume-average relations
(\ref{vol-var3}) and (\ref{Lambda(R)-3}) are superior for 
computational purposes 
because the series involved are finite rather than infinite.
For example, the asymptotic surface-area function $\Lambda(R)$
for the square lattice is plotted in Fig. \ref{square2}
using (\ref{Lambda(R)-3}) with (\ref{inter-v2-2d}) for $1 \le R \le 4$. 
The function is seen to be aperiodic, but fluctuates around an average
value in a bounded fashion. It is noteworthy that the behavior of $\Lambda(R)$
for larger values of $R$ is qualitatively the same. 
Interestingly, the average value of $\Lambda(R)$ over 
this small interval near $R=0$ (as well as other intervals
of the same length) is quite
close to the infinite-interval average value ${\overline
\Lambda}$ \cite{footnote4}.

The average value of the surface-area function $\Lambda(R)$ 
over all $R$, equal to the surface-area coefficient $\overline \Lambda$
[cf. (\ref{S-avg})], is
given (to six significant figures) by
${\overline \Lambda} =0.457649$.
The series (\ref{S-avg})  for the square lattice
was first evaluated by Kendall \cite{Ken48}.
Because it is a slowly converging series, he exploited certain
results of number theory to re-express the sum
in terms of a more rapidly convergent
series. 

\begin{table}[bthp]
\caption{ The surface-area coefficient ${\overline \Lambda}$
for some ordered and disordered two-dimensional
point patterns. For ordered lattices, $\phi$ represents the
close-packed covering fraction.}\vspace{0.05in}
\begin{tabular}{c|c|c} \hline \hline 
Pattern  & $\phi$ &  ${\overline \Lambda}/\phi^{1/2}$
\\ \hline \hline
Triangular Lattice& $\pi/\sqrt{12}\approx 0.9069$ &0.508347\\ \hline 
Square Lattice & $\pi/4 \approx 0.7854 $ &0.516401\\ \hline
Honeycomb Lattice& $\pi/(3\sqrt{3}) \approx 0.6046$&0.567026\\ \hline 
Kagom{\'e} Lattice& $3 \pi/(8\sqrt{3})\approx 0.6802$~& 0.586990 \\ \hline
Step+Delta-Function $g_2$  & 0.5 & $2^{5/2}/(3\pi) \approx 0.600211$\\ \hline 
Step-Function $g_2$& 0.25 & $8/(3\pi)\approx 0.848826$ \\ \hline
One-Component Plasma& $---$&$2/\sqrt{\pi}\approx 1.12838$ \\ \hline \hline
\end{tabular}
\label{2d}
\end{table}

We found that numerical evaluation
of the ensemble-average 
relation (\ref{S-ensemble2}) is a simple and effective means
of computing accurately the surface-area coefficient ${\overline \Lambda}$
for any common lattice.
In two dimensions, this relation yields
\begin{equation}
{\overline \Lambda} = \lim_{ \beta\rightarrow 0^+} 
\left[\frac{16\phi^2}{\pi^{1/2}\beta^{3/2} }- 
\frac{8\phi}{\pi}\sum_{k=1}^{\infty} Z_k r_k e^{\displaystyle-\beta r_k^2}  \right].
\label{sum-2}
\end{equation}
The sum in (\ref{sum-2}) is easily
computed  as a function of the convergence parameter $\beta$
for any simple lattice. For sufficiently small $\beta$, this
sum is linear in $\beta$ and  extrapolation to $\beta \rightarrow 0^+$
yields results that are accurate to at least
six significant figures. We have also computed
the surface-area coefficient for 
triangular, honeycomb, and Kagom{\' e} lattices.
The result for the triangular lattice was first
reported by Kendall and Rankin \cite{Ken53}, but the results
for the honeycomb and Kagom{\' e} lattices are new.
In Table \ref{2d}, we compare all of these results
for the common two-dimensional lattices to one another
by tabulating the normalized scale-independent surface-area
coefficient, i.e., ${\overline \Lambda}/\phi^{1/2}$ [cf.
(\ref{Lambda-scaling})].  
Rankin \cite{Ra53} proved that the triangular lattice
has the smallest normalized
surface-area coefficient for circular windows among all infinite
periodic two-dimensional lattices, which is borne
out in Table \ref{2d}.  However, there is no 
proof that the triangular lattice minimizes ${\overline \Lambda}/\phi^{1/2}$
among all infinite two-dimensional hyperuniform point patterns
for circular windows.
Included in Table \ref{2d} are
results for disordered point patterns that will be discussed in
the ensuing sections.

\begin{figure}[bthp]
\centerline{\psfig{file=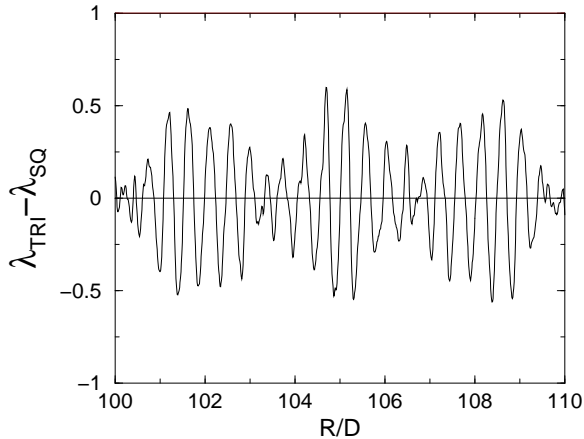,width=3.0in}}
\vspace{-0.2in}
\caption{The difference between the normalized scale-independent
surface-area function $\Lambda(R)$ for the triangular
and square lattices as a function of $R$, where 
$D$ is the lattice spacing. Here $\lambda_{\mbox{tri}}=
\Lambda(R)_{\mbox{tri}}/\phi_{\mbox{tri}}^{1/2}$
and   $\lambda_{\mbox{sq}}=
\Lambda(R)_{\mbox{sq}}/\phi_{\mbox{sq}}^{1/2}$.}
\label{tri-sq}
\end{figure}

Although the normalized surface-area coefficient is smallest
for the triangular lattice, Table \ref{2d} reveals that
the corresponding coefficients for the other lattices 
are not appreciably larger. This suggests that the 
fluctuating surface-area function $\Lambda(R)$ for non-triangular
lattices may be smaller than the corresponding
function for the triangular lattice for certain values of $R$.
This is indeed the case as illustrated in
Figure \ref{tri-sq}, where the difference between
the normalized scale-independent surface-area function $\Lambda(R)$
for the triangular and square lattices is plotted for 
the range $100 D \le R \le 110 D$ using relation (\ref{Lambda(R)-3}).
This difference oscillates rapidly about zero
over this range of $R$, but the same qualitative
trends occur for all values of $R$ and for any pair
of periodic lattices considered here. Given our previous
interpretation of the global minimum of the variance
as corresponding to the ground state of a many-particle
system with a  potential energy function given by $\alpha(r;R)$
(Section III), we see that
the optimal lattice structure is sensitive to
small changes in the value of $R$ (which determines the range of
the potential). This calls into question
previous studies \cite{Ja98} that claim to have found stable 
ground-state lattices for two-dimensional systems
of particles with purely repulsive interaction potentials
of the same qualitative form as shown for $\alpha(r;R)$
in Fig. \ref{intersection} with $d=2$.

\subsection{Three-Dimensional Examples}
\vspace{-0.2in}

Here we specialize to common infinite three-dimensional periodic lattices:
simple cubic (SC) lattice, face-centered cubic (FCC) lattice, hexagonal-close-packed (HCP) lattice,
body-centered cubic (BCC), and the diamond lattice. Explicit
results for the number variance for such lattices have
heretofore not been reported. From  (\ref{parseval}),
(\ref{S-period}) and (\ref{S-avg}) with $d=3$, we have general 
three-dimensional relations for the variance $\sigma^2(R)$, asymptotic surface-area
function $\Lambda(R)$ and surface-area coefficient
${\overline \Lambda}$, respectively. 
These expressions are easily evaluated for the
specific lattice given $v_C$ and the reciprocal lattice vectors $\bf q$.
As we noted earlier,
the volume-average relations (\ref{vol-var3}) and (\ref{Lambda(R)-3}) 
for $d=3$ are superior for computational purposes provided
that $R$ is not very large. Qualitatively, the three-dimensional
trends for the surface-area function $\Lambda(R)$ are similar
to the two-dimensional ones described above (see, for example,
Figs. \ref{square2} and \ref{tri-sq}) and so we will not explicitly
present such three-dimensional results here. 
 
\begin{table}[bthp]
\caption{ The surface-area coefficient ${\overline \Lambda}$
for some ordered and disordered three-dimensional
point patterns. For ordered lattices, $\phi$ represents the
close-packed covering fraction.}\vspace{0.05in}
\begin{tabular}{c|c|c} \hline \hline
Pattern  & $\phi$ & ${\overline \Lambda}/\phi^{2/3} $
\\ \hline \hline
BCC Lattice & $3 \pi/(8\sqrt{3})\approx 0.6802$&1.24476\\ \hline
FCC Lattice &$\pi/\sqrt{18}\approx 0.7405 $  &1.24552\\ \hline
HCP Lattice & $\pi/\sqrt{18} \approx 0.7405$ &1.24569 \\ \hline
SC Lattice & $\pi/6\approx 0.5236$ &1.28920 \\ \hline
Diamond Lattice&$3 \pi/(16\sqrt{3})\approx 0.3801$~ &1.41892 \\ \hline 
Damped-Oscillating $g_2$ & 0.46 &1.44837 \\ \hline
Step+Delta-Function $g_2$  & $0.3125$ & $ 5^{1/3}\cdot  9/2^{10/3} \approx 1.52686$\\ \hline 
Step-Function $g_2$& 0.125 & 2.25 \\ \hline \hline
\end{tabular}
\label{3d}
\end{table}

The ensemble-average relation (\ref{S-ensemble2}),
which for $d=3$ and $D=1$ yields
\begin{equation}
{\overline \Lambda} = \lim_{ \beta\rightarrow 0^+}
\left[\frac{72\phi^2}{\beta^2}- 
6\phi\sum_{k=1} Z_k r_k e^{\displaystyle-\beta r_k^2}  \right],
\label{sum-3}
\end{equation}
and  provides an efficient means
of computing the surface-area coefficient $\overline \Lambda$
for three-dimensional infinite periodic lattices by extrapolating
the results for sufficiently small $\beta$ to $\beta \rightarrow 0^+$
This has been carried out for all of the aforementioned
common three-dimensional lattices and the results
are summarized in Table \ref{3d}, where we  tabulate
the normalized scale-independent surface-area
coefficient, i.e., ${\overline \Lambda}/\phi^{2/3}$ [cf.
(\ref{Lambda-scaling})]. 

Contrary to the expectation
${\overline
\Lambda}/\phi^{d/(d-1)}$
should, among all lattices,
be a global minimum for the closest-packed 
lattices for spherical windows, we find that the  minimum 
in three dimensions is achieved for the BCC lattice, albeit very close in numerical value
to the FCC value (the next smallest value) \cite{footnote5}. This suggests
that the  closest-packed Bravais lattice for $d \ge 3$ does
not minimize ${\overline \Lambda}/\phi^{d/(d-1)}$ \cite{footnote2}.
 Included in Table \ref{3d} are
results for disordered point patterns that will be discussed in
the ensuing sections.


\section{Non-Periodic Hyperuniform Systems}

In this section, we briefly describe the known non-periodic
hyperuniform point patterns in one, two, and three dimensions
and identify some others. For certain one-, two, and three-
dimensional disordered hyperuniform point patterns, we exactly determine 
the corresponding surface-area coefficients, structure factors, direct correlation 
functions, and their associated critical exponents.
A discussion concerning the potential use of 
surface-area coefficient $\overline \Lambda$
as an order metric for general hyperuniform
point patterns is reserved for Section VI.

\subsection{Examples}

Statistically homogeneous hyperuniform point patterns
in one dimension are not difficult to construct. Two
examples are discussed here: one is a ``lattice-gas''
type model and the other is a construction due to 
Goldstein {\it et al.} \cite{Go03}. The first example is constructed
by tesselating the real line into regular intervals of length $D$.
Then a single point is placed in each interval (independently of the others)
at any real position with uniform random 
distribution. The number density $\rho=1/D$,
and the pair correlation function is simply given
by
\begin{equation}
g_2(r) = \left\{\begin{array}{ll}
r/D , \qquad & r\leq D,\nonumber \\
1 , \qquad & r>  D,
\end{array}
\right.
\end{equation}
One can easily verify that the system is hyperuniform
($A=0$)  and that the surface-area coefficient (\ref{S-ensemble}) is given by
\begin{equation}
{\overline \Lambda}=\frac{1}{3},
\end{equation} 
exactly twice the surface-area
coefficient  for the simple single-scale
periodic point pattern [cf. (\ref{1/6})].  This one-dimensional lattice-gas
model is a special case of the so-called $d$-dimensional shuffled
lattice that we will describe below.

A less trivial example of a statistically homogeneous
one-dimensional hyperuniform system is the
construction of Goldstein {\it et al.} \cite{Go03}, 
which obtains from a homogeneous Poisson
point process a new hyperuniform point process.
This construction is defined as follows: First, one
defines a statistically homogeneous process $X(x)$ on the real line such
that $X(x) \le 1$. This process is specified by
dynamics such that $X(x)$ decreases at the rate of unity, except at the
points of the Poisson process, where $X(x)$ jumps up by
one unit unless this jump violates the upper bound condition,
in which case no jump occurs. Second, one takes the points
of the new point process to be those points in which $X(x)$
actually jumps. This new point process is hyperuniform.
It is not known how to extend this construction to higher
dimensions ($d \ge 2$).

The construction of statistically homogeneous and isotropic point patterns 
that are hyperuniform in two or higher dimensions is a challenging 
task. An example of a statistically homogeneous $d$-dimensional system
that is hyperuniform is the so-called {\it shuffled lattice} \cite{Ga03},
but it is not statistically isotropic. This is a lattice
whose sites are independently randomly displaced by a distance
$x$ in all directions according to some distribution with
a finite second moment.

Gabrielli et al. \cite{Ga03} have observed that the
a point pattern derived from the ``pinwheel'' tiling of the plane \cite{Ra94}
has a number variance that grows as the surface-area (perimeter) 
of the window, and is statistically homogeneous and isotropic.
The prototile of the pinwheel tiling 
is a right triangle with sides of length one, two, and $\sqrt{5}$. The 
tiling is generated by performing certain ``decomposition''
and ``inflation'' operations on the prototile. In the first step, the prototile
is subdivided into five copies of itself and then these new triangles 
are expanded to  the size of the original triangle. These decomposition 
and inflation operations are repeated {\it ad infinitum} until the triangles
completely cover the plane (see Fig. \ref{pinwheel}). It is 
obvious from the aforementioned discussion 
that the point pattern that results by randomly placing a point in each 
elementary triangle is hyperuniform. Importantly, because the tiles appear
in {\it infinitely} many orientations, one can show that the resulting
pattern is not only  statistically homogeneous
but statistically isotropic. The full rotational invariance
of the pattern is experimentally manifested by a diffraction
pattern consisting of uniform rings rather than isolated Bragg
peaks.

\begin{figure}[bthp]
\centerline{\psfig{file=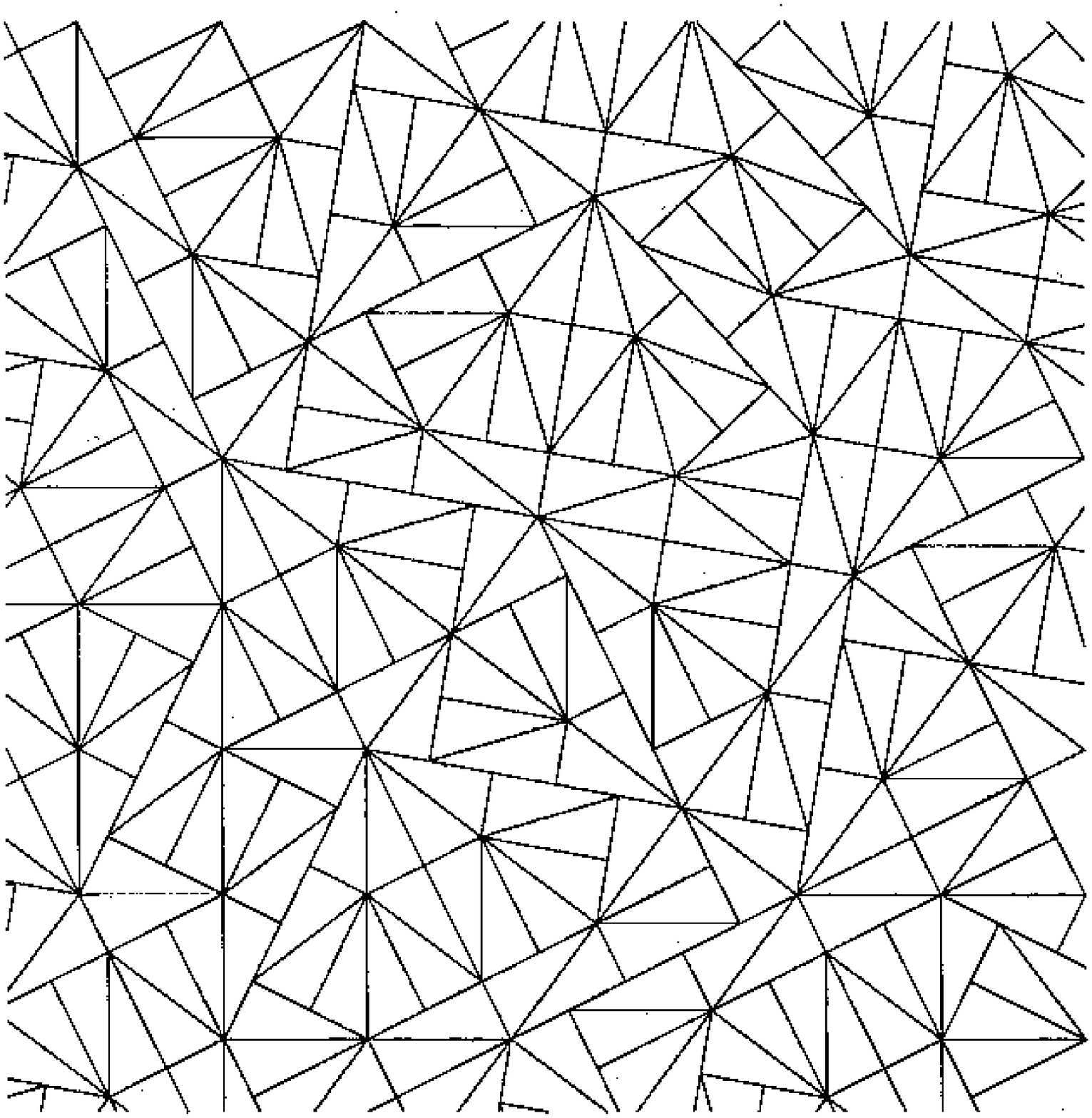,width=3.0in}}
\caption{Portion of a pinwheel tiling.}
\label{pinwheel}
\end{figure}

The one-component plasma is a statistical mechanical
model that is known to have a number variance that
grows only as the surface area of the window \cite{Ma80,Le00}.
The one-component plasma is a system of point particles
of charge $e$ embedded in a uniform background that
imparts overall charge neutrality. In $d=2$, the $n$-particle
correlation functions for this model are exactly solvable
in the thermodynamic limit when the coupling constant
$\Gamma \equiv e^2/(kT)=2$ \cite{Ja80}, and, in particular,
the total correlation function is then given by
\begin{equation}
h(r)=-e^{-\pi\rho r^2}.
\label{ocp}
\end{equation}
Substitution of (\ref{ocp}) into (\ref{S-ensemble2}) gives the surface
area coefficient \cite{Le00} as
\begin{equation}
{\overline \Lambda}=\frac{2}{\sqrt{\pi}}\phi^{1/2},
\end{equation}
where $\phi=\rho \pi D^2/4$. This evaluation of ${\overline \Lambda}$ is
included in Table \ref{2d}. Observe that the structure factor
of the $d$-dimensional one-component plasma at small $k$ behaves as
\begin{equation}
S(k) \sim k^2 \qquad (k \rightarrow 0)
\end{equation}
and, therefore, the corresponding asymptotic behavior of the
Fourier transform of the direct correlation function is
given by
\begin{equation}
c(k) \sim -\frac{1}{k^2} \qquad (k \rightarrow 0).
\end{equation}

Another interesting model that is known to be hyperuniform
\cite{Pi02,Ga03} is the Harrison-Zeldovich \cite{Ha70}
power spectrum for the primordial
density fluctuations in the universe. Here the structure
factor for small $k$ behaves as
\begin{equation}
S(k) \sim k.
\end{equation}
Recently, Gabrielli et al. \cite{Ga03} have discussed the 
construction of point patterns in three dimensions
that are consistent with the Harrison-Zeldovich spectrum.

The present authors have recently introduced 
and studied so-called $g_2$-invariant
processes \cite{St01,Sa02,To02b}. A $g_2$-invariant process
is one in which a chosen nonnegative form for
the pair correlation function $g_2$ remains
invariant over a nonvanishing density range while keeping
all other relevant macroscopic variables fixed. The upper
limiting ``terminal'' density is the point above which
the nonnegativity condition on the structure factor
[cf. (\ref{S-k})] would be violated. Thus, at the terminal
or critical density, the system is hyperuniform if realizable. In the
subsequent subsection, we will calculate the surface-area
coefficient exactly for several of these $g_2$-invariant
processes. We will also exactly determine the corresponding structure factors,
direct correlation functions,  and 
their associated critical exponents.

Interestingly, random packings of spheres near the maximally random
jammed (MRJ) state \cite{To00,Ka02} appear to be hyperuniform.  
Figure \ref{MRJ} depicts the structure factor for such
a computer-generated 40,000-particle packing, 
is vanishingly small for small wavenumbers.  The packing is
strictly jammed \cite{To01}, which means that the particle system
remains mechanically rigid under attempted global 
deformations (including shear) that do not increase volume
and, furthermore the packing is saturated. A {\it saturated} 
packing of hard spheres is one in which there is no space available to add another 
sphere. In the case of saturated packings of identical 
hard spheres of unit diameter, 
no point in space has distance greater than unity from the center of some
sphere. An attractive postulate would be that all strictly jammed saturated 
infinite packings of identical spheres are hyperuniform. 
Examples of strictly jammed saturated periodic packings in two and three dimensions
include the closest packed triangular and face-centered cubic
lattices, respectively. In light of this discussion, one can view
a disordered packing near the MRJ state as a type of ``glass"
for the hard-sphere system. An important open fundamental
question is whether there are molecular glasses
(with ``soft'' intermolecular potentials) that become
hyperuniform in the limit that the temperature vanishes. Indeed,
our preliminary results indicate that this possibility
is attainable.

\begin{figure}[bthp]
\centerline{\psfig{file=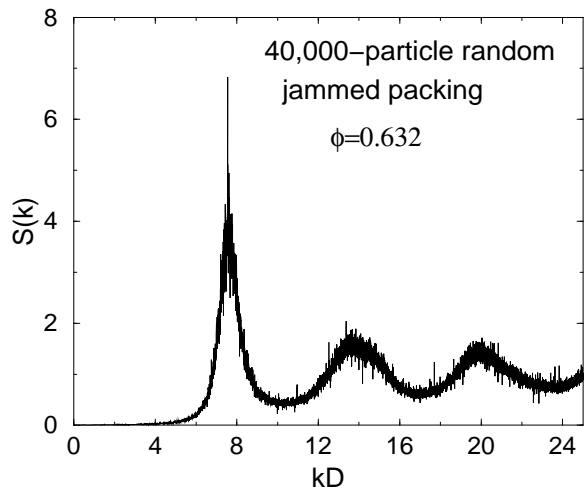,width=3.0in}}
\caption{The structure factor for a random packing of three-dimensional
identical hard spheres of diameter $D$ near the MRJ state \protect\cite{To00,Ka02}
as computed from a single realization
consisting of 40,000 particles in a cubical box with periodic
boundary conditions using the
protocol described in Ref. \cite{Ka02}. 
The packing (covering) fraction of spheres
$\phi$ is 0.632. }
\label{MRJ}
\end{figure}

\subsection{Exact Results for $g_2$-Invariant Processes}

 Here we evaluate the surface-area
 coefficient exactly for three different disordered
 $g_2$-invariant processes studied by us earlier \cite{St01,Sa02,To02b}. 
 We also exactly determine the corresponding
 structure factors, direct correlation functions,  and 
 their associated critical exponents.

\subsubsection{Step-Function $g_2$}

Let us first consider the $g_2$-invariant process in which
a spherically symmetric pair correlation or radial
distribution function is defined 
by the unit step function \cite{St01}
\begin{equation}
g_2(r)=\Theta(r-D)=\left\{\begin{array}{ll}
0, \qquad & r\leq D, \nonumber \\
1, \qquad & r>D,\\
\end{array}
\right.
\label{step}
\end{equation}
The condition $g_2(r)=0$ for $r \le D$ prevents any pair
of points from getting closer than a distance $D$ to
one another. Note that in the special case of a system of 
identical hard spheres in equilibrium in the limit $\rho \rightarrow 0$,
$g_2$ is exactly given by (\ref{step}). The corresponding total
correlation function is given by
\begin{equation}
h(r)=-\Theta(D-r)=\left\{\begin{array}{ll}
-1, \qquad & r\leq D, \nonumber \\
0, \qquad & r>D,\\
\end{array}
\right.
\label{step-h}
\end{equation}
which when substituted into (\ref{A2}) and (\ref{B2})
yields the volume and surface-area coefficients as
\begin{equation}
A=S(k=0)=1-2^d\phi, \qquad B=\frac{\overline \Lambda}{2^d \phi}=
\frac{ 2^{d-2}d^2\Gamma(d/2)}
{\Gamma((d+3)/2)
\Gamma(1/2)} \phi.
\end{equation}
The reduced density $\phi$ defined by (\ref{phi}) 
(equivalent to the covering fraction of the hard cores
of diameter $D$) lies
in the range $0 \le \phi \le \phi_c$, where 
\begin{equation}
\phi_c=\frac{1}{2^d}
\end{equation}
is the terminal or critical density, i.e., the density
at which the system is hyperuniform, where $A=0$ and 
\begin{equation}
B={\overline \Lambda}=
\frac{d^2 \Gamma(d/2)}
{4\Gamma((d+3)/2)
\Gamma(1/2)}.
\label{lambda-1}
\end{equation}
The values of the scale-independent surface-area
coefficient ${\overline \Lambda}/\phi^{(d-1)/d}$ for $d=1,2$ and 3 are given
in Tables II, III and IV, respectively.
It is noteworthy that a recent study \cite{Cr03} provides
convincing numerical evidence
that the step-function $g_2$ is realizable by systems of impenetrable
$d$-dimensional spheres (with $d=1$ and $d=2$)
for densities up to the terminal density. 
Thus,  it appears that satisfying the nonnegativity conditions 
on $g_2(r)$ and $S(k)$ in this instance is sufficient to ensure realizability. 

The Fourier transform of the total correlation function
(\ref{step-h})  yields the analytic function
\begin{equation}
{\tilde h}(k)=-\left(\frac{2\pi}{kD}\right)^{d/2} D^d J_{d/2}(kD).
\end{equation}
Thus, use of (\ref{factor}) gives the structure factor for $\phi$ in the
range $0 \le \phi \le \phi_c$ to be
\begin{equation}
S(k)=1-\Gamma(1+d/2) 
\left(\frac{2}{kD}\right)^{d/2}
\left(\frac{\phi}{\phi_c}\right) J_{d/2}(kD).
\end{equation}
Similarly, the Ornstein-Zernike relation (\ref{c}) yields an
exact expression for the Fourier transform of the direct
correlation function:
\begin{equation}
{\tilde c}(k)= \frac{\displaystyle -\left(\frac{2\pi}{kD}\right)^{d/2} D^d
J_{d/2}(kD)}{\displaystyle 1-\Gamma(1+d/2) 
\left(\frac{2}{kD}\right)^{d/2}
\left(\frac{\phi}{\phi_c}\right) J_{d/2}(kD)}.
\end{equation}

Thus, the small-$k$ expansions of 
$S(k)$ and ${\tilde c}(k)$, which determine their behavior
in the vicinity of, and at, the critical point,   are respectively given by
\begin{equation}
S(k)=\displaystyle \left(1-\frac{\phi}{\phi_c} \right)
+ \frac{1}{2(d+2)} \frac{\phi}{\phi_c}(kD)^2 +{\cal O}[(kD)^4]
\label{S-step}
\end{equation}
and 
\begin{equation}
{\tilde c}(k)=\frac{-v_1(D) }{\displaystyle 
 \left(1-\frac{\phi}{\phi_c} \right)
+ \frac{1}{2(d+2)} \frac{\phi}{\phi_c}(kD)^2 +{\cal O}[(kD)^4]},
\label{c-step}
\end{equation}
where $v_1(D)$ is the volume of a $d$-dimensional sphere
of radius $D$ [cf. (\ref{v1})].
At the critical point $\phi=\phi_c$, we see that $S(k) \sim k^2$
and ${\tilde c}(k) \sim - k^{-2}$, and therefore comparison
to (\ref{S-eta}) and (\ref{c-eta}) yields the exponent $\eta=0$.
Relation (\ref{S-step}) leads to the power law 
\begin{equation}
S^{-1}(0)=\left(1-\frac{\phi}{\phi_c} \right)^{-1}, \qquad \phi
\rightarrow \phi_c^{-}, 
\end{equation}
which upon comparison to (\ref{S-power}) immediately
yields the critical exponent $\gamma=1$. The correlation length
$\xi$ is defined via (\ref{c-step}), which we rewrite as
\begin{equation}
k^2 {\tilde c}(k)+ \xi^{-2} {\tilde c}(k)= -G, \qquad k D \ll 1
\label{k-diff1}
\end{equation}
where
\begin{eqnarray}
\xi &=& \frac{D}{[ 2(d+2)\phi_c]^{1/2}} \left(1-\frac{\phi}{\phi_c} \right)^{-1/2}, 
\qquad \phi \rightarrow \phi_c^{-}, \label{xi-step}\\ \nonumber \\
G &=& \frac{2(d+2) v_1(D)}{D^2} \frac{\phi_c}{\phi}, \label{G-step}
\end{eqnarray}
and $v_1(D)$ is the volume of a sphere of radius $D$
defined by (\ref{v1}). Comparison of (\ref{xi-step}) to
the power law (\ref{c-power}) yields the exponent $\nu =1/2$.
Note that the exponent values $\gamma=1$, $\xi=1/2$,
and $\eta=0$ are consistent with the interrelation (\ref{inter}).
Inversion of (\ref{k-diff1}) yields the partial differential
equation
\begin{equation}
\nabla^2 c(r)-\xi^{-2} c(r)= G \delta({\bf r}), \qquad r \gg D,
\label{diff1}
\end{equation}
where the spherically symmetric Laplacian operator $\nabla^2$ in any dimension $d$ is given by 
\begin{equation}
\nabla^2=\frac{1}{r^{d-1}} \frac{\partial}{\partial r} \left[ r^{d-1} \frac{\partial }
{\partial r}\right].
\label{Laplace}
\end{equation}
We see that the direct correlation function in real space for large $r$ 
is determined by the Green's function of the linearized Poisson-Boltzmann equation.

Let us first determine the solutions of (\ref{diff1})  
at the critical point $\phi=\phi_c$ where $\xi$ diverges to
infinity. Thus, the asymptotic behavior of $c(r)$ for $r \gg D$ is given by the infinite-space Green's function for the $d$-dimensional Laplace equation \cite{To00}, and so we obtain
\begin{equation}
c(r) = \cases{
\displaystyle{-6 \left(\frac{r}{D}\right)}, 
     &$d = 1$, \cr
 \displaystyle{4 \ln \left(\frac{r}{D}\right)}, 
     &$d = 2$, \cr
 \displaystyle{- \frac{2(d+2)}{d(d-2) } \left(\frac{r}{D}\right)^{d-2}},     
     &$d \ge 3$.
}
\label{c-green1}
\end{equation}
Observe that it is only for $d \ge 3$ that $c(r)$ follows the power-law
form (\ref{c-eta}) with an exponent
$\eta=0$. The fact that $\eta$ takes on an integer value is due to the fact
that ${\tilde h}(k)$ is an analytic function.
Note also that the real-space direct correlation function of 
the one-component plasma has precisely the same
asymptotic form as (\ref{c-green1}), albeit with different
amplitudes (prefactors).

As $\xi \rightarrow \infty$  for fixed
$r$, the solutions of (\ref{diff1}) are
\begin{equation}
c(r) = \cases{
\displaystyle{-6 \frac{\phi_c}{\phi} \left(\frac{\xi}{D}\right) \exp(-r/\xi)}, 
     &$d = 1$, \cr
 \displaystyle{4 \frac{\phi_c}{\phi}\ln \left(\frac{r}{D}\right) \exp(-r/\xi)}, 
     &$d = 2$, \cr
 \displaystyle{- \frac{2(d+2)\phi_c}{d(d-2) \phi}
 \left(\frac{r}{D}\right)^{d-2}\exp(-r/\xi)},     
     &$d \ge 3$.
}
\label{c-green2}
\end{equation}
On the other hand, it is noteworthy that as $r \rightarrow \infty$
for fixed $\xi$, the asymptotic behavior changes according to the relation
\begin{equation}
c(r)=- \frac{(d+2)\sqrt{2\pi}\phi_c}{\Gamma(1+d/2) \phi}
\left(\frac{D}{\xi}\right)^{(d-3)/2} \left(\frac{D}{r}\right)^{(d-1)/2}
\exp(-r/\xi), \qquad d \ge 1.
\end{equation}

\subsubsection{Step+Delta Function $g_2$}

Here we consider the $g_2$-invariant process defined by
a radial distribution function  that consists 
of the  aforementioned unit step function plus a delta function contribution
that acts at $r=D$:
\begin{equation}
g_2(r)=\Theta(r-D)+ \frac{Z}{\rho s_1(D)}\delta(r-D), 
\label{step-delta}
\end{equation}
where $Z$ is a nonnegative constant and $s_1(D)$ is
the surface area of a sphere of radius $D$ defined by (\ref{area-sph}).
The function (\ref{step-delta}) was one of several examples studied by Torquato
and Stillinger \cite{To02b} to understand the relationship
between short-range order and maximal density in sphere
packings. In this investigation, $Z$ was interpreted
as the average contact coordination number. Here we consider
their case IV (given in the appendix of Ref. \cite{To02b}) in which
the condition
\begin{equation}
Z=\frac{2^d d}{d+2}\phi
\end{equation}
is obeyed in order to constrain the location of the minimum
of the structure factor to be at $k=0$. Here the reduced density $\phi$ lies
in the range $0 \le \phi \le \phi_c$, and
\begin{equation}
\phi_c=\frac{d+2}{2^{d+1}}
\end{equation}
is the terminal or critical density. Note that the function 
specified by relation (\ref{step-delta}) is a special limit
of the radial distribution function corresponding to the 
dilute and narrow 
limit of the square-well potential studied by Sakai, Stillinger,
and Torquato \cite{Sa02}.

Substitution of (\ref{step-delta}) into (\ref{A2}) and (\ref{B2})
yields the volume and surface-area coefficients as
\begin{equation}
A=S(k=0)=1-\frac{2^{d+1}}{d+2}\phi, \qquad B=\frac{\overline \Lambda}{2^d \phi}=
\frac{ 2^{d-2}d^2 \Gamma(d/2)}
{(d+2)\Gamma((d+3)/2)
\Gamma(1/2)}\phi.
\end{equation}
At the critical density, $A=0$ and 
\begin{equation}
{\overline \Lambda}=2^d \phi_c B=\frac{ d^2 
(d+2)\Gamma(d/2)}
{16\Gamma((d+3)/2)
\Gamma(1/2)}.
\label{lambda-2}
\end{equation}
The values of the scale-independent surface-area
coefficient ${\overline \Lambda}/\phi^{(d-1)/d}$ for $d=1,2$ and 3 are given
in Tables \ref{1d}, \ref{2d} and \ref{3d}, respectively.

The combination of  relations (\ref{factor}), (\ref{c}), and (\ref{step-delta}) give the structure factor and Fourier transform of the direct
correlation function respectively for $\phi$ in the
range $0 \le \phi \le \phi_c$:
\begin{equation}
S(k)=1+
\frac{2^{d/2} \Gamma(2+d/2) }{(kD)^{(d/2)-1}}
\left(\frac{\phi}{\phi_c}\right) \Bigg[\frac{J_{(d/2)-1}(kD)}{d+2}-\frac{J_{d/2}(kD)}{kD}\Bigg],
\end{equation}
\begin{equation}
{\tilde c}(k)= \frac{\displaystyle \frac{(2\pi)^{d/2} D^d }{(kD)^{(d/2)-1}}
 \Bigg[\frac{\displaystyle J_{(d/2)-1}(kD)}{d+2}-
\frac{\displaystyle J_{d/2}(kD)}{kD}\Bigg]}
{\displaystyle 1+
\frac{2^{d/2} \Gamma(2+d/2) }{(kD)^{(d/2)-1}}
\left(\frac{\phi}{\phi_c}\right) \Bigg[\frac{J_{(d/2)-1}(kD)}{d+2}-
\frac{J_{d/2}(kD)}{kD}\Bigg]}
\end{equation}
Therefore, the Taylor expansions of 
$S(k)$ and ${\tilde c}(k)$ about $k=0$ are respectively given by
\begin{equation}
S(k)=\displaystyle \left(1-\frac{\phi}{\phi_c} \right)
+ \frac{1}{8(d+2)(d+4)} \frac{\phi}{\phi_c}(kD)^4 +{\cal O}[(kD)^6]
\label{S-step-delta}
\end{equation}
and 
\begin{equation}
{\tilde c}(k)=\frac{-2v_1(D) }{\displaystyle 
 \left(1-\frac{\phi}{\phi_c} \right)
+\frac{1}{8(d+2)(d+4)} \frac{\phi}{\phi_c}(kD)^4 +{\cal O}[(kD)^6]}.
\label{c-step-delta}
\end{equation}

Relation (\ref{S-step-delta}) leads to the power law 
\begin{equation}
S^{-1}(0)=\left(1-\frac{\phi}{\phi_c} \right)^{-1}, \qquad \phi
\rightarrow \phi_c^{-}, 
\end{equation}
which upon comparison to (\ref{S-power}) again
yields the critical exponent $\gamma=1$. The correlation length
$\xi$ is defined via (\ref{c-step-delta}), which we rewrite as
\begin{equation}
k^4 {\tilde c}(k)+ \xi^{-4} {\tilde c}(k)= -G, \qquad k D \ll 1
\label{k-diff2}
\end{equation}
where
\begin{eqnarray}
\xi &=& \frac{D}{[ 8(d+2)(d+4)\phi_c]^{1/4}} \left(1-\frac{\phi}{\phi_c} \right)^{-1/4}, 
\qquad \phi \rightarrow \phi_c^{-}, \label{xi-step-delta}\\ \nonumber\\
G &=& \frac{16(d+2)(d+4)v_1(D)}{D^4} \frac{\phi_c}{\phi} . \label{G-step-delta}
\end{eqnarray}
Comparison of (\ref{xi-step-delta}) to
the power law (\ref{c-power}) yields the exponent $\nu =1/4$.
We see that the exponent values $\gamma=1$, $\xi=1/4$,
and $\eta=-2$ are consistent with the interrelation (\ref{inter}).
Inversion of (\ref{k-diff2}) yields the partial differential
equation
\begin{equation}
\nabla^4 c(r)+\xi^{-4} c(r)= -G \delta({\bf r}), \qquad r \gg D,
\label{diff2}
\end{equation}
where $\nabla^4 \equiv \nabla^2\nabla^2$ is the spherically symmetric 
biharmonic operator, and $\nabla^2$ is given by (\ref{Laplace}). 

The solutions of (\ref{diff2})  
at the critical point $\phi=\phi_c$ ($\xi \rightarrow \infty$)
are given by the infinite-space
Green's function for the $d$-dimensional biharmonic equation. It is only for
$d \ge 5$ that the solutions admit a power law of the form (\ref{c-power})
with an exponent $\eta=-2$, namely,
\begin{equation}
c(r) = - \frac{8(d+2)(d+4)}{d(d-2)(d-4) }\left(\frac{D}{r}\right)^{d-4}, \qquad d \ge 5.
\label{c-green3}
\end{equation}

\subsubsection{Damped-Oscillating $g_2$}

In three dimensions, Torquato and Stillinger \cite{To02b}
also considered a $g_2$-invariant process that appends a
damped-oscillating contribution to the aforementioned
step+delta function $g_2$. Specifically, they examined
the radial distribution function
\begin{equation}
g_2(r)= \Theta(r-D)+ \frac{Z}{\rho 4\pi D^2}\delta(r-D)+
\frac{a_1}{r}e^{-a_2r} \sin(a_3 r+a_4)\Theta(r-D).
\end{equation}
Here we consider their case II, where at the terminal
density $\phi_c=0.46$, $Z=2.3964$, $a_1=1.15$, $a_2=0.510$,
$a_3=5.90$, and $a_4=1.66$. At this critical point,
the volume coefficient $A=0$ and the surface-area coefficient
(\ref{S-ensemble}) is given by
\begin{equation}
{\overline \Lambda}=36\phi_c^2-6\phi_c Z+144 a_1\phi_c^2 I,
\label{oscill}
\end{equation}
where
\begin{eqnarray*}
I &=& \int_1^{\infty}x e^{-a_2 x} \sin(a_3 x+a_4) dx\nonumber
\\
&=& \frac{(2a_3^3-a_3^5-6a_2^2a_3-2a_2^2a_3^3-4a_2^3a_3
-4a_2a_3^3-a_2^4a_3)}{(a_2^2+a_3^2)^3}e^{-a_2}\cos(a_3+a_4) \nonumber\\
&& + \frac{(2a_3^4+6a_2a_3^2-a_2^5-2a_2^4-2a_2^3-a_2a_3^4
-2a_2^3a_3^2)}{(a_2^2+a_3^2)^3}e^{-a_2}\sin(a_3+a_4). 
\end{eqnarray*}
Substitution of the  aforementioned parameters in (\ref{oscill}) yields
${\overline \Lambda}=0.863082$. This evaluation of $\overline \Lambda$ 
is included in Table \ref{3d}.
With this choice of $g_2$, the first non-zero term
of the small-$k$ expansion of the 
structure factor $S(k)$ at the critical point is
of order $k^4$, and therefore the exponent $\eta=-2$,
as in the previous case. However, here $c(r)$ does not
admit the power-law form (50) for large
$r$ because $\eta < -1$.

\section{Discussion and Conclusions}

     The principal theme presented in this paper is that number fluctuations 
calculated for variable window geometries offer a powerful tool to characterize 
and to classify point-particle media.  This theme encompasses both spatially 
periodic (crystalline) particle patterns, as well as those that are globally 
disordered (amorphous).  By considering the large-window asymptotic limit, 
special attention attaches to "volume" and to "surface" fluctuations in 
space dimension $d \ge 1$.  
A special class of ``hyperuniform" point patterns has 
been recognized for which the volume fluctuations vanish identically; 
equivalently these are systems for which the structure factor
$S(k)$ vanishes at $k=0$.  
Another special class of ``hyposurficial" point patterns has 
also been recognized for which the surface fluctuations vanish identically.  The first of these 
special attributes requires that the ($d-1$)-st spatial moment of the total 
correlation function be constrained in magnitude; the second requires a 
similar constraint on the $d$-th spatial moment of the total correlation 
function.  The preceding text demonstrates that no point pattern can 
simultaneously be both hyperuniform and hyposurficial. 
     
     All infinitely extended perfectly periodic structures are hyperuniform.  
We have stressed that geometrically less regular cases of hyperuniformity also 
exist, including those that are spatially uniform and isotropic.  The suitably 
normalized surface fluctuation quantity, which measures the extent to which 
hyperuniform systems fail to attain hyposurficial status, becomes a natural 
nonnegative order metric that we have evaluated numerically for a basic 
sampling of structures.  We proved that the 
simple periodic linear array yields the global minimum value
for hyperuniform patterns in $d=1$,
and showed that the triangular lattice  
produces the smallest values for the cases 
tested in $d=2$.  But in spite of the fact that these 
minimizing structures correspond to optimal packings of rods and disks, 
respectively, the face-centered-cubic lattice for optimal sphere packing 
does not minimize the surface-fluctuation order metric for $d=3$.  Instead, the 
body-centered cubic lattice enjoys this distinction \cite{footnote3}.  For each choice of 
space dimension, other lattices and irregular hyperuniform patterns yield 
higher values for this order metric. An order metric for hyperuniform
systems based on the local variance may find potential use
in categorizing ``jammed'' and ``saturated'' sphere packings
\cite{To00,Ka02,To01,Ka02b}
whose long-wavelength density fluctuations vanish.
     
     It is clearly desirable to extend the set of point patterns for which the 
surface fluctuation order metric has been numerically evaluated.  This would 
help to strengthen the impression created thus far that regardless of space 
dimension $d$, point patterns arranged by increasing values of the order metric 
are indeed essentially arranged by increasing structural disorder.  It will 
be important in the future to include a selection of two and three-dimensional 
quasicrystalline point patterns \cite{St86} in the comparisons; the 
presumption at the present state of understanding is that they would present 
order metrics with values that lie between the low magnitudes of periodic 
lattices, and the substantially larger magnitudes of spatially uniform, 
isotropic, irregular point patterns.  It would also benefit insight to include 
cases of spatially uniform, but anisotropic, point patterns;
for example,  those 
associated with ``hexatic" order in two dimensions \cite{KT}.
     
     An important class of hyperuniform systems arises from the so-called 
``$g_2$-invariant processes" \cite{To02b,Cr03,St01,Sa02}.  These 
processes 
require that the pair correlation function $g_2(r)$ remain unchanged as density increases 
from zero. For those $g_2$-invariant processes
that correspond to thermal equilibrium, this criterion is
implemented by virtue of compensating continuous changes in the particle pair 
potential function.  For any given choice of the invariant $g_2$, such a process is 
in fact achievable, but only for densities up to a terminal density limit.  
At this upper limit, the system of points attains hyperuniformity, i.e., 
$S(k)=0$.
Furthermore, examination of the Ornstein-Zernike relation reveals that the 
direct correlation function $c(r)$ 
develops a long-range tail as the terminal density 
is approached from below.  By implication, for the special
case of a thermal equilibrium process, the pair potential at the terminal 
density develops a long-range repulsive Coulombic form.  The conclusion is 
that hyperuniformity at that terminal density is logically associated with 
the local electroneutrality condition that all equilibrium systems of 
electrostatically charged particles must obey \cite{St68}.
     
     The Ornstein-Zernike relation, though originally conceived to apply to 
systems in thermal equilibrium, can nevertheless be formally applied to any 
system for which the pair correlation function $g_2(r)$
is available.  Hyperuniform 
systems that are irregular and isotropic possess short-range pair correlation 
only, but as in the examples just cited the corresponding direct correlation 
functions are long-ranged.  In an important sense, hyperuniform systems exhibit 
a kind of ``inverted critical phenomenon".  For conventional 
liquid-vapor critical 
points, $h(r) \equiv g_2(r)-1$ is long-ranged and 
implies diverging density fluctuations and isothermal 
compressibilities, while the direct correlation function $c(r)$
remains short-ranged.  
Hyperuniform systems have short range for $h(r)$, vanishing volume fluctuations and 
isothermal compressibility, and a long-ranged $c(r)$.
     
     As a final matter, we mention that an attractive direction for future 
study of hyperuniformity and related concepts involves consideration of 
collective density variables.  These are defined by a nonlinear transformation 
of point-particle positions ${\bf r}_j$  ($1 \le j \le N$) as follows:
\begin{equation}
\rho({\bf k})=\sum_{j=1}^N \exp(i {\bf k} \cdot {\bf r}_j).
\end{equation}
If the particles interact through a spherically-symmetric pair potential 
whose Fourier transform exists and is denoted by $V(k)$, then the overall potential 
energy for the $N$ particles in volume $V$ 
can be expressed in the following manner:
\begin{equation}
\Phi=\frac{1}{2V} \sum_{\bf k} V(k)[\rho({\bf k})\rho(-{\bf k})-N].
\end{equation}
It has been demonstrated \cite{Fa91}
that at least in one dimension, application of a suitable $V(k)$, followed
by $\Phi$ 
minimization, can totally suppress density fluctuations for $\bf k$'s near the origin.  
This automatically produces a hyperuniform system configuration.  
Analogous studies need to be pursued for two- and three-dimensional systems.

\begin{acknowledgments}
We are grateful to Joel Lebowitz for bringing
to our attention the problem of surface fluctuations
and for many insightful
discussions.  We thank Aleksandar Donev, Andrea Gabrielli, Hajime
Sakai, and Peter Sarnak for very helpful comments.
This work was supported by the Petroleum Research Fund 
as administered by the American
Chemical Society, MRSEC Grant at Princeton University, NSF
DMR-0213706, and by the Division of Mathematical Sciences at
NSF undewr Grant No. DMS-0312067.
\end{acknowledgments}

\begin{appendix}

\section{Intersection Volume of Two Identical $d$-Dimensional Spheres}

In this appendix, we obtain an explicit expression for the
scaled intersection volume of two identical
$d$-dimensional spheres of radius $R$ whose centers
are separated by a distance $r$. This function $\alpha(r;R)$
is defined by (\ref{I}).

We begin by noting that the $d$-dimensional Fourier transform
(\ref{fourier}) of any integrable function $f(r)$ that  depends only on
the modulus $r=|\bf r|$ of the vector $\bf r$ is
given by \cite{Sn95}
\begin{equation}
{\tilde f}(k) =\left(2\pi\right)^{\frac{d}{2}}\int_{0}^{\infty}r^{d-1}f(r)
\frac{J_{\left(d/2\right)-1}\!\left(kr\right)}{\left(kr\right)^{\left(d/2\right
)-1}}dr,
\end{equation}
and the inverse transform (\ref{fourier-inverse}) of $f(k)$ is given by
\begin{equation}
f(r) =\frac{1}{\left(2\pi\right)^{\frac{d}{2}}}\int_{0}^{\infty}k^{d-1}f(k)
\frac{J_{\left(d/2\right)-1}\!\left(kr\right)}{\left(kr\right)^{\left(d/2\right
)-1}}dk.
\label{inverse}
\end{equation}
Here $k$ is the modulus of the wave vector $\bf k$ 
and $J_{\nu}(x)$ is the Bessel function of order $\nu$.

The Fourier transform of the window indicator function (\ref{window2}) is given by
\begin{eqnarray}
{\tilde w}(k;R) &=& \frac{(2\pi)^{d/2}}{k^{(d/2)-1}}\int_0^R
r^{d/2} J_{(d/2)-1}(kr) dr \nonumber \\
&=& \left(\frac{2\pi}{kR}\right)^{d/2} R^d J_{d/2}(kR).
\label{window-Fourier}
\end{eqnarray}
Therefore, the Fourier transform of $\alpha(r;R)$, defined by (\ref{I-f}),
is given by
\begin{equation}
{\tilde \alpha}(k;R)= 2^d \pi^{d/2} \Gamma(1+d/2)\frac{[J_{d/2}(kR)]^2}{k^d}.
\end{equation}
Using the inverse transform (\ref{inverse}) yields
the scaled intersection volume function to be
\begin{eqnarray}
\alpha(r;R) &=& \frac{2^d \Gamma(1+d/2)}{r^{(d-2)/2}} \int_0^\infty
\frac{[J_{d/2}(kR)]^2 J_{\left(d/2\right)-1}(kr)} {k^{d/2}} dk \nonumber \\
&=& I_{1-x^2}\left(\frac{d+1}{2},\frac{1}{2}\right) \Theta(2R-r),
\label{I2}
\end{eqnarray}
where 
\begin{equation}
I_x(a,b)=\frac{B_x(a,b)}{B(a,b)}
\end{equation}
is the {\it normalized} incomplete beta function \cite{Ab72},
\begin{equation}
B_x(a,b)=\int_0^x t^{a-1}(1-t)^{b-1}dt,
\end{equation}
is the incomplete beta function, and
\begin{equation}
B(a,b)=\int_0^1 t^{a-1}(1-t)^{b-1}dt=\frac{\Gamma(a)\Gamma(b)}{\Gamma(a+b)}
\end{equation}
is the beta function.

For the first five space dimensions, relation (\ref{I2}), for $r \le 2R$, yields
\begin{eqnarray}
\alpha(r;R)& = &  1 -\frac{r}{2R}
\label{inter-v2-1d}, \qquad d = 1, \\
\alpha(r;R) & = &  \frac{2}{\pi}
\left[ \cos^{-1}\left(\frac{r}{2R}\right) - \frac{r}{2R} 
\left(1 - \frac{r^2}{4R^2}\right)^{1/2} \right], \qquad d = 2,
\label{inter-v2-2d}\\
\alpha(r;R) & = & 1 -\frac{3}{4} \frac{r}{R}+ \frac{1}{16}\left(\frac{r}{R}\right)^3 ,
\qquad d = 3, \label{inter-v2-3d} \\
\alpha(r;R)& = & \frac{2}{\pi}
\left[ \cos^{-1}\left(\frac{r}{2R}\right) - \left\{\frac{5r}{6R}-
\frac{1}{12}\left(\frac{r}{R}\right)^3\right\} 
(1 - \frac{r^2}{4R^2})^{1/2} \right], \quad d = 4,\\
\alpha(r;R) & = & 1 -\frac{15}{16} \frac{r}{R}+ \frac{5}{32}\left(\frac{r}{R}\right)^3 -\frac{3}{256}\left(\frac{r}{R}\right)^5, \qquad d=5.
\end{eqnarray}
Figure \ref{intersection} shows graphs of the scaled intersection
volume $\alpha(r;R)$ as a function of $r$ for the first
five space dimensions. For any dimension, $\alpha(r;R)$
is a monotonically decreasing function of $r$. At a fixed
value of $r$ in the open interval $(0,2R)$, $\alpha(r;R)$
is a monotonically decreasing function of the dimension $d$.

\begin{figure}[bthp]
\centerline{\psfig{file=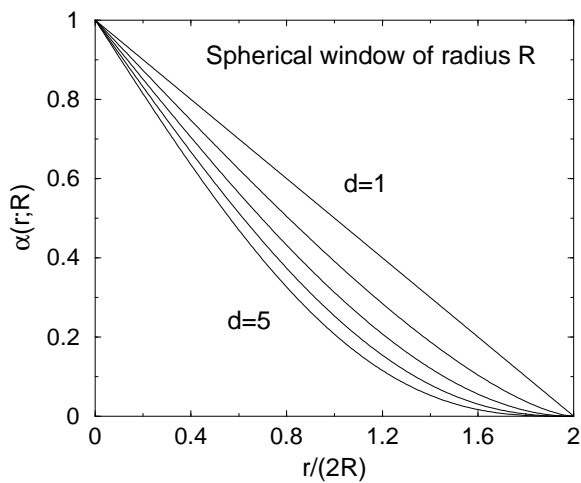,width=3.0in}}
\caption{The scaled intersection
volume $\alpha(r;R)$ for spherical windows of radius $R$
as a function of $r$ for the first
five space dimensions. The uppermost curve is for $d=1$
and lowermost curve is for $d=5$.}
\label{intersection}
\end{figure}

Expanding the general expression (\ref{I2}) through first
order in $r$ for $r \le 2R$ yields
\begin{equation}
\alpha(r;R)= 1 -\frac{\Gamma(\frac{d}{2}+1)}{\Gamma(\frac{d+1}{2})
\Gamma(\frac{1}{2})}\frac{r}{R} + {o}\left(\frac{r}{R}\right),
\label{alpha}
\end{equation}
where $o(x)$ indicates terms of higher order than $x$.
This relation will be of use to us in developing an asymptotic
expression for the number variance for large windows.

\section{Fluctuations in Equilibrium Hard-Particle Systems}

Hard particles in equilibrium  represent an example of a correlated
system that is generally not hyperuniform. The one-dimensional
case of identical hard rods of length $D$ in equilibrium is
a particularly instructive case because the radial distribution
function $g_2(r)$ (in the thermodynamic limit) is known exactly 
for all densities \cite{Ze27}:
\begin{equation}
\phi g_2 (x) = \sum_{k = 1}^\infty \; \Theta(x - k) \frac{\phi^{k}(x - k)^{k-1}}
{(1-\phi)^{k}
(k - 1)!} \exp\left[ - \frac{\phi(x - k)}{1-\phi}\right],
\label{g2-1d}
\end{equation}
where $x = r/D$ is a dimensionless distance and $\phi=\rho D$
is the covering fraction of the rods, which lies
in the closed interval $[0,1]$. Below the close-packed space-filling
value of $\phi=1$, the radial distribution
function is a short-ranged function in the sense
that one can always find a large enough value of $r$ beyond
which $g_2(r)$ remains appreciably close to unity. 
That is, for $\phi <1$, the correlation length
is always finite. However, the point $\phi=1$ is singular
in the sense that the system exhibits perfect long-range
order and thus is hyperuniform.
Indeed, at $\phi=1$, the nearest-neighbor distance
for each rod is exactly equal to $D$: a situation that is identically the
same as the single-scale one-dimensional periodic point pattern
studied in Section IV.

\begin{figure}[bthp]
\centerline{\psfig{file=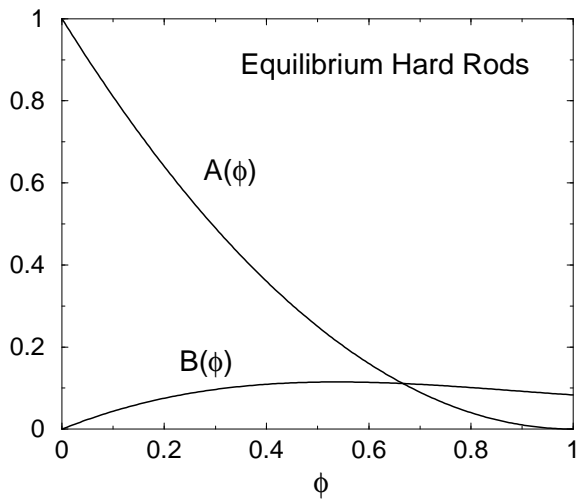,width=3.0in}}
\vspace{-0.2in}
\caption{ The volume coefficient $A(\phi)=(1-\phi)^2$ and surface-area
coefficient $B(\phi)$ [defined in relation (\ref{var1})] as
a function of the reduced density $\phi$ for a one-dimensional
system of identical hard rods in equilibrium. At
the hyperuniform density $\phi=1$,
$B={\overline \Lambda}/2=1/12$, which
corresponds to the perfectly ordered close-packed state.} 
\label{rods}
\end{figure}

Using relation (\ref{g2-1d}) in conjunction with relations (\ref{A1}) and
(\ref{B1}) enables us to compute the ``volume'' and ``surface-area'' 
contributions to the variance as a function of reduced density $\phi$ for 
identical hard rods in equilibrium. The results are summarized in Fig. \ref{rods}.
We see that as the density increases, the volume fluctuations
decrease monotonically and only vanish at the space-filling density
$\phi=1$: the hyperuniform state. Of course, $B$ vanishes at $\phi=0$
and increases in value as $\phi$ increases until it achieves
a maximum value at $\phi \approx 0.5$. At the hyperuniform 
state ($\phi=1$), $B={\overline \Lambda}/2=1/12$, which
corresponds to the perfectly ordered close-packed state.
For sufficiently small densities, the surface-area coefficient 
of equilibrium hard-sphere systems in higher dimensions is expected
to have the same qualitative behavior as the one-dimensional
case. Specifically, the same trends should occur in higher dimensions 
for densities in the range $0 \le \phi \le \phi_f$, where $\phi_f$ 
corresponds to the freezing density, i.e., the point above which the 
system undergoes a disorder to order phase transition. For
densities between freezing and melting points, the behavior
of the surface-area coefficient is expected to be qualitatively
different from that for hard rods in equilibrium, which
is devoid of a phase transition. However, we can definitively
assert that the highest achievable density along the stable
crystal branch is a hyperuniform state. In particular,
for hard disks ($d=2$) and hard spheres ($d=3$) in equilibrium,
the hyperuniform states correspond to the close-packed
triangular lattice and 
FCC lattice, respectively.

\section{How Small Can the Volume Coefficient be for
Hyposurficial Systems?}

We know that a statistically homogeneous
and isotropic point pattern cannot simultaneously be
hyperuniform and hyposurficial, i.e.,  the
volume coefficient $A$ [cf. (\ref{A2})] and 
surface-area coefficient $B$ [cf. (\ref{B2})]
cannot both be zero for a strictly convex window (Section IIC).
The purpose of this appendix is to investigate
how small $A$ can be made for an infinite  hyposurficial point pattern ($B=0$).
To that end we consider a hypothetical  spherically symmetric
pair correlation function $g_2(r)$ and a spherical window.
We do not place any additional restrictions on $g_2(r)$
besides the necessary realizability conditions that $g_2(r) \ge 0$
for all $r$ and $S(k) \ge 0$ for all $k$. 
The hypothetical correlation function is characterized by  three
parameters $\epsilon$, $C$,  and $D$ as follows: 
\begin{equation}
g_2(r)=g_S(r)+g_L(r),
\label{hypoth}
\end{equation}
where $g_S(r)$ denotes the short-ranged part defined
by the step function
\begin{equation}
g_{S}(r) = \left\{\begin{array}{ll}
 0, \qquad & 0 \leq r \leq D, \\
1 , \qquad & r > D,
\end{array}
\right.
\label{short}
\end{equation}
and $g_L(r)$ denotes the long-ranged part defined by
\begin{equation}
g_{L}(r) = \left\{\begin{array}{ll}
0, \qquad & 0 \leq r \leq D, \\
\displaystyle C \epsilon \left(\frac{D}{r}\right)^{d+1+\epsilon} , \qquad & r > D,
\end{array}
\right.
\label{long}
\end{equation}
Here $D$ is a length
parameter, $C$ is a dimensionless
constant, and $\epsilon$
is a positive ($\epsilon  > 0$) but small parameter.
The necessary condition $g_2(r) \ge 0$ requires
that the constant $C$ satisfy the trivial inequality
\begin{equation}
C \ge -\epsilon^{-1} .
\label{ineq}
\end{equation}

The form of $g_2$ ensures that we can make the surface-area coefficient
$B$ vanish identically, as required. According to relation (\ref{B2}), the surface-area coefficient
$B$ is proportional to the $d$th moment of the total correlation
function $h(r)=g_2(r)-1$. The $d$th moment integral for
the hypothetical pair correlation function (\ref{hypoth}) is given by
\begin{equation}
\int_0^{\infty} h(r) r^d dr=-\frac{D^{d+1}}{d+1}+ CD^{d+1}.
\end{equation}
To make this integral vanish, we take 
\begin{equation}
C=\frac{1}{d+1} > 0,
\end{equation}
which of course satisfies the inequality (\ref{ineq}).
For such a hyposurficial correlation function (\ref{hypoth})
that also satisfies the nonnegativity condition $S(k) \ge 0$, we 
now show that the volume coefficient $A$ is only nonzero by ${\cal O}(\epsilon^2)$.

Consider volume coefficient $A$ [cf. (\ref{A3})] with this value of $C$:
\begin{equation}
A=\lim_{|{\bf k}| \rightarrow 0}S({\bf k})=1+\rho\int_{\Re^d} 
h({\bf r}) d{\bf r}=1-2^d \phi +\left(\frac{2^d d \phi}{d+1}\right)\frac{\epsilon}
{1+\epsilon},
\end{equation}
where $\phi=\rho v_1(D/2)$ is a dimensionless density.
If one incorrectly sets $A$ to be zero, one finds that
the corresponding density is given by
\begin{equation}
\phi_*=\frac{1}{\displaystyle 2^d\left(1-\frac{d \epsilon}{(d+1)(1+\epsilon)}\right)}.
\label{phic}
\end{equation}
At such a value of $\phi$, however, $S(k)$ will be negative
for some $k >0$ near the origin for sufficiently
small but nonzero $\epsilon$, which shows in this specific instance
that the point pattern corresponding to such a hypothetical
$g_2$ cannot simultaneously be hyperuniform and hyposurficial,
as expected. However, one can make $S(k=0)$ positive and very small
(while satisfying $S(k) \ge 0$ for all $k$) at a value
$\phi$ slightly smaller than (\ref{phic}) in the limit
$\epsilon \rightarrow 0^+$.

The other necessary condition $S(k) \ge 0$ will  be obeyed
for all $k$ provided that the number density is no larger than some
``terminal density'' $\rho_c$ (or $\phi_c$) \cite{St01,Sa02,To02b}. The structure
factor is given by
\begin{equation}
S(k)=1+ \rho[H_S(k)+H_L(k)],
\end{equation}
where $H_S(k)$ and $H_L(k)$ are the Fourier transforms
of $g_S(r)-1$ and $g_L(r)$, respectively. The terminal
density is given by 
\begin{equation}
\rho_c=-\frac{1}{\min_{k} [H_S(k)+H_L(k)]}.
\end{equation}

For simplicity, we will specialize to the case $d=3$, keeping in
mind that our general conclusions apply to arbitrary dimension.
Based on the aforementioned arguments, it is sufficient 
to consider the behavior of $S(k)$ for small $k$:
\begin{eqnarray}
S(k)&=& 1+8\phi\left[-1+\frac{(kD)^2}{10}-{\cal O}[(kD)^4]\right]
+6\phi\frac{\epsilon}{1+\epsilon} \nonumber \\
&&-\frac{3 \phi\sqrt{\pi}\epsilon}{2^{1+\epsilon}(1+\epsilon)}
\frac{\Gamma(\frac{1}{2}-\frac{\epsilon}{2})}
{\Gamma(2+\frac{\epsilon}{2})}|kD|^{1+\epsilon}
+\frac{\phi\epsilon}{1-\epsilon}(kD)^2+{\cal O}[(kD)^4].
\end{eqnarray}
The nonanalytic term $|kD|^{1+\epsilon}$
[which arises due to inclusion
of relation (\ref{long}) for $g_L$]
has the effect of displacing
the minimum of $S(k)$ away from the origin when $g_L=0$ to a
symmetric pair of locations determined by
\begin{equation}
|k_{min}D| =\frac{15\pi}{16} \epsilon
\end{equation}
as $\epsilon \rightarrow 0^+$. Moreover, in this leading order
\begin{equation}
S(0)-S(k_{min})= \frac{45\pi^2\phi}{64} \epsilon^2.
\end{equation}
Note that this would lead to an ${\cal O}(\epsilon^2)$
correction to expression (\ref{phic}) for $\phi_*$.
In summary, by adopting the correlation function
(\ref{hypoth}) with $C=1/(d+1)$, we can make the
surface-area coefficient $B=0$ and at the terminal
density $\phi_c$, the structure factor $S(0)=A$
is only nonzero by ${\cal O}(\epsilon^2)$.

\end{appendix}

~

\begin{thebibliography}{20}
\bibitem{note}
Of course, there may be situations in which $A=0$, but $B$
and/or higher-order moments of $h$ are unbounded. In such
instances, the Fourier transform of $h$ will be a nonanalytic
function of $k$ at $k=0$.
\bibitem{footnote4}
Indeed, this is true for other periodic lattices
and for statistically homogeneous point patterns 
in two and three dimensions. The fact that the asymptotic
surface-area coefficient $\overline \Lambda$
can be estimated from the average of $\Lambda(R)$
over a small range of $R$ near $R=0$ can be exploited
to speed up significantly optimization algorithms
to find extremal or targeted values of $\overline \Lambda$.
\bibitem{footnote5}
A proof that the extremal lattice for $d=3$ is the BCC lattice
currently does not exist. In this connection, it should be
noted that the surface-area coefficient for the BCC lattice
is related to the Epstein zeta function for the FCC lattice,
which was shown [V. Ennola, {\it Proc. Camb. Phil. Soc.}
{\bf 60}, 855 (1964)] to provide a local minimum value
among all lattices.
\bibitem{footnote3}
It is an open question whether the triangular lattice
and the body-centered cubic lattice correspond
to the global minimum value of ${\overline \Lambda}/\phi^{(d-1)/d}$
among all infinite hyperuniform point patterns
in two and three dimensions, respectively, for spherical
windows.
\end{thebibliography}
\end{document}